\documentclass[twocolumn]{aastex63}

\usepackage{bm}

\newcommand{\zd}{Z_{\rm d}}
\newcommand{\tg}{T_{\rm g}}
\newcommand{\nh}{n_{\rm H}}
\newcommand{\kb}{k_{\rm B}}
\newcommand{\md}{m_{\rm d}}
\newcommand{\sigd}{\sigma_{\rm d}}
\newcommand{\jpe}{J_{\rm pe}}
\newcommand{\jion}{J_{\rm ion}}
\newcommand{\je}{J_{\rm e}}
\newcommand{\jsec}{J_{\rm sec,gas}}
\newcommand{\qabs}{Q_{\rm abs}}
\newcommand{\qsca}{Q_{\rm sca}}
\newcommand{\qpr}{Q_{\rm pr}}

\newcommand{\ype}{Y_{\rm pe}}
\newcommand{\fdrag}{F_{\rm drag}}
\newcommand{\frad}{F_{\rm rad}}
\newcommand{\fgrav}{F_{\rm grav}}

\newcommand{\br}{\bm{r}}
\newcommand{\bvp}{\bm{v}_{\rm p}}
\newcommand{\brp}{\bm{r}_{\rm p}}
\newcommand{\vp}{v_{\rm p}}
\newcommand{\rp}{r_{\rm p}}
\newcommand{\phip}{\phi_{\rm p}}
\newcommand{\zp}{z_{\rm p}}
\newcommand{\vrel}{v_{\rm rel}}
\newcommand{\rinit}{r_{\rm init}}
\newcommand{\vterm}{v_{\rm term}}

\newcommand{\tacc}{t_{\rm acc}}
\newcommand{\tspt}{t_{\rm sp}}
\newcommand{\tcoag}{t_{\rm coag}}

\newcommand{\unitnum}{\rm{cm}^{-3}}
\newcommand{\unitb}{\rm{erg}~\rm{cm}^{-3}}

\shorttitle{Dust Destruction by Drift-Induced Sputtering in AGN}
\shortauthors{Tazaki and Ichikawa}

\begin{document}

\title{Dust Destruction by Drift-Induced Sputtering in Active Galactic Nuclei}

\correspondingauthor{Ryo Tazaki}
\email{rtazaki@astr.tohoku.ac.jp}

\author[0000-0003-1451-6836]{Ryo Tazaki}
\affil{Astronomical Institute, Graduate School of Science
Tohoku University, 6-3 Aramaki, Aoba-ku, Sendai 980-8578, Japan}

\author[0000-0002-4377-903X]{Kohei Ichikawa}
\affil{Astronomical Institute, Graduate School of Science
Tohoku University, 6-3 Aramaki, Aoba-ku, Sendai 980-8578, Japan}
\affil{Frontier Research Institute for Interdisciplinary Sciences, Tohoku University, Sendai 980-8578, Japan}




\begin{abstract}
Recent mid-infrared high spatial resolution observations have revealed that active galactic nuclei (AGNs) may host a polar dust region with the size of several pc, and such dust may be carried by radiation from the central engine. 
The polar dust emission often exhibits 
very weak or absence of the silicate 10-$\mu$m emission feature. A possible explanation is that the polar dust is dominated by micron-sized large grains because these grains do not show the silicate feature,
while it remains unclear how large grains are preferentially supplied to the polar region.
We here propose a new scenario describing the prevalence of large grains at the polar region. We show that grains are accelerated to the hypersonic drift velocity by the radiation pressure
from AGN, and the hypersonic drift results in dust destruction via kinetic sputtering. Sputtering destroys small grains faster than
the large ones, and thus larger grains will be preferentially blown over longer distance. Although the hypersonic drift, or kinetic sputtering, tends to be suppressed for very small grains due to the Coulomb drag, they might also be disrupted by Coulomb explosion.
Removal of small grains and/or survival of large grains may explain the lack of a silicate 10-$\mu$m emission feature in polar dust emission.
\end{abstract}



\section{Introduction}\label{sec:intro}

Dust is a crucial component of active galactic nuclei (AGNs). 
Mid-infrared (MIR) emission is ubiquitously observed in AGNs, 
and the emission is always compact 
\citep[$<10$~pc; e.g.,][]{pac05,hon10,ram11,alo11,asm14,ich15}. This suggests that there is a compact dusty region heated by optical/ultraviolet (UV) emission from the central engine that is re-emitted thermally in the MIR \citep[e.g.,][]{gan09,ich12,asm15,ich17a,ich19a}.

Recent MIR high spatial resolution interferometric observations have
partially resolved the compact dusty region of nearby AGNs ($D<50$~Mpc); these regions have a size of a few pc \citep[e.g.,][]{bur13}. Interestingly, these dusty regions appear to be elongated in the polar direction \citep{hon12,hon13,tri14,lop16,lef18,hon19}, as opposed to the equatorial direction where the conventional dusty region, commonly termed as ``torus'', is considered to be located \citep[e.g.,][]{urr95}.

The above observations have prompted discussions on the origin of the polar-elongated dusty region.
Phenomenological models have been developed by several
groups \citep{hon17,sta17}; the main conclusion from these studies 
is that two components are necessary to successfully reproduce the resolved near-IR to MIR spectral energy distributions (SEDs), specifically, a geometrically thin dusty disk in the equatorial direction and a hollow, cone-like polar dusty region launching from the inner dust sublimation region of the disk.

To reproduce such geometrical structures, more physically motivated radiative hydro-dynamical simulations have been incorporated \citep[e.g.,][]{kro07,cha16,wil18}.
Notably, \cite{wad15} proposed that a radiation-driven fountain model could reproduce such a polar dusty region.
In this model, the dust is no longer static, but rather a dynamically moving failed dusty outflow or ``fountain'' that produces the geometrically thick polar-elongated structure. The fountain is supported by the radiation pressure from the geometrically thin gas/dust disk.

One notable observational feature of the polar dust emission is that  silicate 10-$\mu$m emission feature is very weak or sometimes absent \citep[][]{hon12,hon13,bur13,hon19}, especially when compared to the unresolved single dish telescope data. In order to explain such weak silicate feature, sublimation of small silicate grains or dusty outflow composed of graphite grains have been discussed \citep[e.g.,][]{hon12}.
Another possibility for weak silicate feature is that MIR emission from the polar region is optically thick \citep[e.g.,][]{lao93} because if dusty-gas is optically thick, its emission spectrum approaches to the black body.

Here, we propose a new possibility, where the polar dust is predominantly composed of micron-sized or larger grains, since large silicate grains do not efficiently show silicate feature \citep[e.g.,][]{lao93}. In this case, dusty-gas at the polar region is not necessary to be optically thick.  In order to suppress silicate feature, sub-micron-sized grains should be removed. Coulomb explosion could eliminate very small grains ($\lesssim0.1~\mu$m) \citep{WDB06,RT20a}; however, it is insufficient to destroy sub-micron-sized grains ($\gtrsim0.1~\mu$m) at the pc-scale regions. Hence, it is not evident how large grains are preferentially supplied to the polar region, while sub-micron-sized grains are removed. 

In this paper, we resolve this problem by considering drift-induced sputtering. We show that dust grains irradiated by the harsh AGN radiation, e.g., in the polar region, are subjected to the hypersonic drift due to radiation pressure, which causes sputtering of grains. Since smaller grains are destroyed faster than large grains \citep[e.g.,][]{D79}, large grains are preferentially blown over larger distance (e.g., several-pc scales).

This paper is organized as follows. In Section \ref{sec:model}, we summarize our AGN and torus models, as well as the basic equations.
In Section \ref{sec:results}, we present results of numerical calculations. A discussion and summary of our findings are presented in Sections \ref{sec:discussion} and \ref{sec:conclusion}, respectively.

\section{Models and Methods} \label{sec:model}
We study two-dimensional motion of a dust grain as well as its destruction irradiated by radiation from the central engine and the dusty torus. Radiation models are given in Section \ref{sec:rad}, and the equation of motion of a dust grain is described in Section \ref{sec:eom}.
Since grain charge is necessary to compute the gas drag force and the sputtering rate, we present a method to compute grain charge in Section \ref{sec:charge}. Evolution of grain radius by sputtering is given in Section \ref{sec:spt}.
For the sake of simplicity, uniform gas is assumed, that is, density and temperature are constant in the calculation. In addition, we assume that gas is at rest.

We adopted a simple model of the dusty disk of
AGNs, which is conventionally referred to as an AGN ``torus''. 
In this paper, cylindrical coordinates $(r,\phi,z)$ were used, with the origin at the central AGN location. 

\subsection{AGN and Dusty Disk Model} \label{sec:rad}
\begin{figure}[t]
\begin{center}
\includegraphics[height=6.0cm,keepaspectratio]{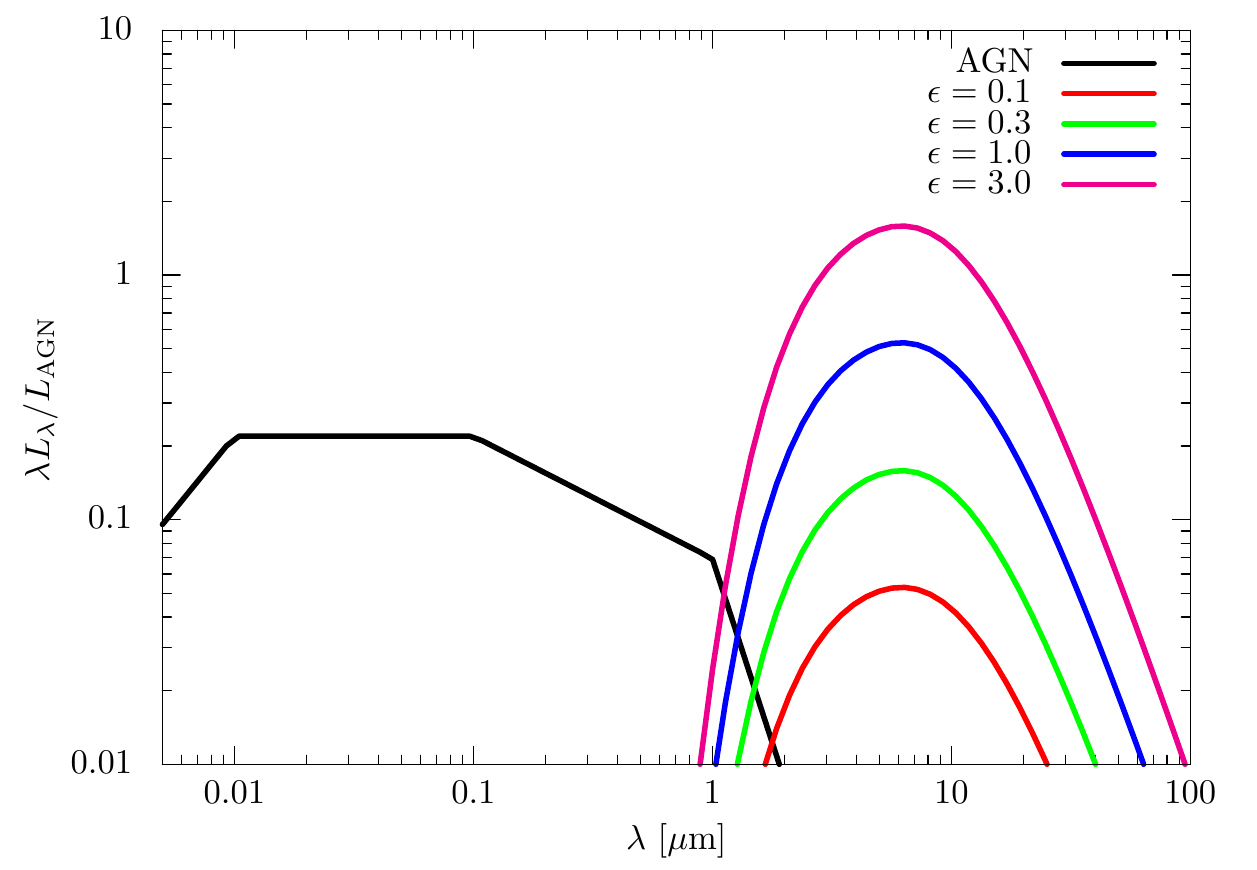}
\caption{Spectral energy distribution (SED) of our model of the active galactic nucleus (AGN) and dusty disk. Ultraviolet (UV) and infrared (IR) components correspond to AGN and dusty disk emission, respectively. For 
the dusty disk model, luminosity models $\epsilon=0.1, 0.3, 1.0,$ and 3.0 are shown.}
\label{fig:sed}
\end{center}
\end{figure}

We adopted the radiation spectra of AGNs used in \citet{N08}
whose parameter is their bolometric luminosity ($L_\mathrm{AGN}$).
As a dusty torus, we assumed a flat dust disk geometry (torus component is confined into $z=0$ plane) for simplicity; this assumption is also in good agreement with recent results suggesting a geometrically thin dust disk, as discussed in Section~\ref{sec:intro}.
Hereafter, we refer to this dusty torus component as an 
AGN ``dusty disk''.
The radial temperature distribution of the dusty disk is assumed to obey a single power law, $T_\mathrm{d}(r)=T_\mathrm{in}(r/r_\mathrm{in})^q$, where $r$ is the radius and $T_\mathrm{in}$ and $r_\mathrm{in}$ are the dust sublimation temperature and radius, respectively. The dust sublimation temperature and power-law index were set as $T_\mathrm{in}=1500$ K and $q=-0.5$. With the AGN bolometric luminosity $L_\mathrm{AGN}=10^{45}$ erg s$^{-1}$, the inner and outer radii of the torus are $r_\mathrm{in}=0.4$ pc \citep{B87, K11, K14} and $r_\mathrm{out}=$10 pc \citep{K11, G16, I16}, respectively.
In this paper, we adopted $M_\mathrm{BH}=10^8M_\odot$,
which yields the Eddington ratio of $\lambda_\mathrm{Edd}\simeq0.08$ 
for $L_\mathrm{AGN}=10^{45}$~erg~s$^{-1}$.
The SED of the dusty disk is given by 
\begin{equation}
\frac{\lambda L_\lambda}{L_\mathrm{AGN}}=\epsilon\frac{8\pi^2\lambda}{L_\mathrm{AGN}}\int_{r_\mathrm{in}}^{r_\mathrm{out}} B_\lambda(T_\mathrm{d}(r))rdr,
\end{equation}
where we have introduced the parameter $\epsilon$, the ratio of the luminosities of the dusty disk and AGN; $\epsilon$ is treated as a free parameter. Figure \ref{fig:sed} presents the SED of our AGN and dusty disk model for various values of $\epsilon$. 
In terms of energy budget, it is expected $\epsilon<1$ because energy source of torus emission is radiation from the central engine. Although $\epsilon>1$ seems to be unrealistic, we also consider such cases as a numerical experiment to study the impact of torus radiation on grain dynamics.

\subsection{Equation of Motion} \label{sec:eom}
We study dynamics of a dust grain at the vicinity of AGN by solving the equation of motion:
\begin{eqnarray}
\frac{d\brp}{dt}&=&\bvp, \label{eq:eom1}\\
\md\frac{d\bvp}{dt}&=&
\bm{F}_\mathrm{grav}
+\bm{F}_\mathrm{rad}
+\bm{F}_\mathrm{rad,torus}
+\bm{F}_\mathrm{drag}, \label{eq:eom2}
\end{eqnarray}
where $\md$ is the mass of a dust grain, $\brp$ and $\bvp$ are the position and velocity vectors of the dust grain, $\bm{F}_\mathrm{grav}$ is the gravitational force due to the central blackhole, $\bm{F}_\mathrm{rad}$ and $\bm{F}_\mathrm{rad,torus}$ are the radiation pressure forces from AGN and a dusty torus, respectively, and $\bm{F}_\mathrm{drag}$ is the gas drag force. 
Initially, the dust grain is located at $r_\mathrm{init}$ away from AGN at rest ($\bm{v}_p=0$).

\subsubsection{Gravity and Radiation Pressure}
The gravitational force acting on a grain due to the central blackhole is
\begin{eqnarray}
\bm{F}_\mathrm{grav}=-\frac{GM_\mathrm{BH}\md}{\rp^2}\frac{\brp}{\rp},
\end{eqnarray}
where $G$ is the gravitational constant, $M_\mathrm{BH}$ is the blackhole mass, and $\rp=|\brp|$ is the distance between AGN and the grain.

The radiation pressure force acting on a grain is
\begin{equation}
\bm{F}_\mathrm{rad}=F_\mathrm{rad}\frac{\brp}{\rp},~\frad=\sigd\int_0^\infty d\nu u_\nu\qpr(\nu), 
\end{equation}
where $\sigd=\pi a^2$ is the geometrical cross section of the dust grain with radius $a$, $\nu$ is frequency of radiation, $u_\nu$ is the specific energy density of radiation, and $\qpr=\qabs+(1-g)\qsca$ is the radiation pressure efficiency; where $\qabs$, $\qsca$, and $g$ are the absorption and scattering efficiencies and the asymmetry parameter, respectively \citep{B83}. $\frad$ can be reduced to more simple expression:
\begin{eqnarray}
\frad=\sigd \frac{L_\mathrm{AGN}}{4\pi \rp^2c}\langle \qpr \rangle_\mathrm{AGN}, \label{eq:frad}
\end{eqnarray}
where $c$ is the speed of light and $\langle Q_\mathrm{pr} \rangle_\mathrm{AGN}$ is the spectrum-averaged radiation pressure efficiency.

The ratio of the radiation pressure from AGN radiation and gravity, $\beta_\mathrm{AGN}$, is given by the following:
\begin{equation}
\beta_\mathrm{AGN}\equiv\frac{\frad}{\fgrav}=\lambda_\mathrm{Edd}\left(\frac{m_\mathrm{p}}{\sigma_\mathrm{T}}\right)\left(\frac{\sigma_\mathrm{d}}{m_\mathrm{d}}
\right)\langle \qpr \rangle_\mathrm{AGN}, \label{eq:betaAGN}
\end{equation}
where $m_\mathrm{p}$ is the proton mass, $\sigma_\mathrm{T}$ is the Thomson cross-section, and $\lambda_\mathrm{edd}$ is the Eddington ratio of the radiation.
Both radiation pressure and gravity are inversely proportional to the square of the distance from the AGN; the ratio, $\beta_{\mathrm{AGN}}$, does not depend on the distance. 

\begin{figure}[tb]
\begin{center}
\includegraphics[height=6.0cm,keepaspectratio]{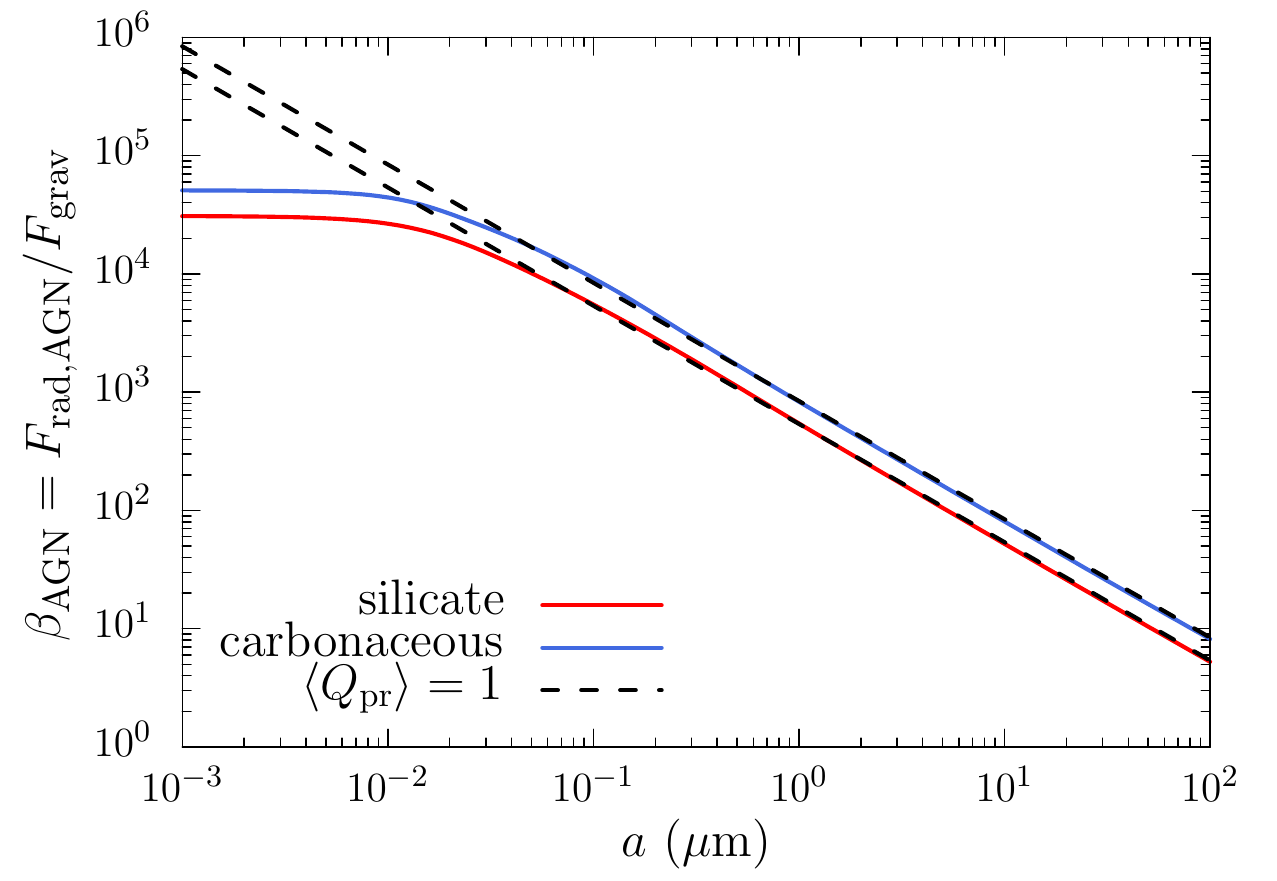}
\caption{The ratio of radiation pressure force to gravity. Red and blue lines show silicate and carbonaceous grains, respectively. Dashed lines show the $\beta$-values for $\langle Q_\mathrm{pr} \rangle_\mathrm{AGN}=1$. The Eddington ratio is set as $\lambda_\mathrm{Edd}=0.1$.}
\label{fig:beta}
\end{center}
\end{figure}

When the grain radius is sufficiently larger than the incident wavelength, we can expect $\langle Q_\mathrm{pr} \rangle_\mathrm{AGN}\simeq1$ (geometrical optics). Substituting $\langle Q_\mathrm{pr} \rangle_\mathrm{AGN}=1$ into Equation (\ref{eq:betaAGN}), we obtain a simple relationship for the radiation pressure force with respect to the gravity:
\begin{equation}
\beta_\mathrm{AGN}\simeq 54\left(\frac{\lambda_\mathrm{edd}}{0.01} \right) \left(\frac{a}{1\ \mu\mathrm{m}}\right)^{-1}\left(\frac{\rho_\mathrm{s}}{3.5\ \mathrm{g}\ \mathrm{cm}^{-3}}\right)^{-1}, \label{eq:beta}
\end{equation}
where $\rho_\mathrm{s}$ is the material density. We assume 
$\rho_\mathrm{s}=3.5~\mathrm{g}~\mathrm{cm}^{-3}$ for silicate and $2.24~\mathrm{g}~\mathrm{cm}^{-3}$ for graphite.

We computed $Q_\mathrm{pr}$ using the Mie theory \citep{B83}, and the optical constant of silicate and graphite grains was taken from \citet{D84, lao93, D03}. 
If the size parameter $x=2\pi{a}/\lambda$, where $\lambda$ is the wavelength, is larger than $2\times10^{4}$, we use the anomalous diffraction approximation \citep{VDH57} instead of using the Mie theory and assume $\qpr\simeq\qabs$. 

Figure \ref{fig:beta} shows the $\beta$-values for silicate and graphite grains with various grain radii.  For $a\gtrsim\mathrm{a~few}\times10^{-2}~\mu$m, Equation (\ref{eq:beta}) well reproduces the $\beta_\mathrm{AGN}$-values.
Since graphite grains have slightly lower material density, they show slightly larger $\beta$-values than silicate grains. A dust grain will be gravitationally unbound if $\beta_\mathrm{AGN}>0.5$ \citep{B79}. Since the $\beta$-values are much larger than 0.5, dust grains will be blown out by AGN radiation pressure as long as other forces are negligible.

\subsubsection{Radiation Pressure from Dusty Torus}
In order to calculate the radiation pressure force due to dusty torus emission, we divide the flat dusty torus by a number of patches.
Let the position vector of a patch center is $\bm{r}$ and each patch has a width of ($\Delta{r},\Delta\phi$), the radiation pressure from each patch may be written as
\begin{equation}
\Delta \bm{F}_\mathrm{rad,torus}=\frac{\sigma_\mathrm{d}}{c}\Delta\Omega \frac{(\brp-\br)}{|\brp-\br|} \epsilon \int_0^\infty  B_\nu(T_\mathrm{d})Q_\mathrm{pr}(a,\nu)d\nu, \label{eq:frad1}
\end{equation}
where $B_\nu$ is the Planck function, $T_\mathrm{d}$ is the temperature of the dust grains, and $\Delta\Omega$ is the solid angle of the patch.
The solid angle is given by
$\Delta\Omega=r\Delta{r}\Delta\phi\cos\theta/d^2$,
where $d=|\brp-\br|$ is the distance between the torus patch and a grain, and $\theta$ is the angle between the unit vector along the $z$-axis (polar direction) and $\brp-\br$. Using the particle coordinate $\brp=(\rp,\phip,\zp)$ and the position vector of a patch center $\br=(r,\phi,0)$, we obtain $\cos\theta=\zp/d$.
Integrating Equation (\ref{eq:frad1}) from $r_\mathrm{in}\le r \le r_\mathrm{out}$ and $0\le\phi\le2\pi$, the radiation pressure due to torus emission becomes
\begin{equation}
\bm{F}_\mathrm{rad,torus}=\frac{\sigma_\mathrm{SB}\sigd}{\pi c}\epsilon\int_{r_\mathrm{in}}^{r_\mathrm{out}}rdr\int_0^{2\pi}d\phi \cos\theta T_\mathrm{d}^4 \langle Q_\mathrm{pr}\rangle_{T_\mathrm{d}}\frac{(\bm{r}_p-\bm{r})}{d^3},
\end{equation}
where $\langle Q_\mathrm{pr}\rangle_{T_\mathrm{d}}$ is the Planck mean radiation pressure efficiency.
 
\subsubsection{Gas Drag} \label{sec:fdrag}
A charged grain moving through ionized gas experiences two kinds of gas drag: (i) direct collision and (ii) Coulomb interaction.
Since we assume that gas is at rest, the velocity of a dust grain $\vp=|\bvp|$ coincides with the relative velocity between gas and dust $\vrel$.
When a dust grain charged with $\zd$ (in electron charge unit) is moving though gas with the relative velocity $v_\mathrm{rel}$, the drag forces can be expressed as \citep{D79}
\begin{eqnarray}
\bm{F}_\mathrm{drag}&=&-F_\mathrm{drag}\frac{\bvp}{\vp}, \\
F_\mathrm{drag}&=&2\sigd k_\mathrm{B}T_\mathrm{g}n_\mathrm{H}\left\{\sum_i A_i [G_0(s_i)+z_i^2\phi^2\ln(\Lambda/z_i)G_2(s_i)\right\}, \label{eq:gasdrag}\nonumber\\ \label{eq:drag} 
\end{eqnarray}
where $k_\mathrm{B}$ is Boltzmann's constant, $\tg$ is the gas temperature, $n_\mathrm{H}$ is the number density of hydrogen atoms, $A_i$ is the abundance of gas species $i$, $z_i$ is the charge of gas species $i$ (in electron charge unit), and $s_i=(m_iv_\mathrm{rel}^2/2k_\mathrm{B}\tg)^{1/2}$ is the normalized relative velocity; where $m_i$ is the mass of gas particles of species $i$. The normalized grain potential $\phi$ and the Coulomb logarithm $\Lambda$ are given by
\begin{eqnarray}
\phi&=&\frac{eU}{\kb\tg},\\
\Lambda&=&\frac{3}{2ae\phi}\left(\frac{\kb\tg}{\pi n_e}\right)^{1/2},
\end{eqnarray}
where $U=e\zd/a$ is the electrostatic potential of a grain and $n_e$ is the number density of electrons.
At the right-hand side of Equation (\ref{eq:drag}), the first and the second terms in the parenthesis represent drag forces due to direct collision and Coulomb interaction, respectively. 
Approximate forms for $G_0$ and $G_2$ are \citep{D79}
\begin{eqnarray}
G_0(s) &\approx&\frac{8s}{3\sqrt{\pi}}\left(1+\frac{9\pi}{64}s^2\right)^{1/2} \label{eq:g0},\\
G_2(s)&\approx& s\left(\frac{3}{4}\sqrt{\pi}+s^3\right)^{-1}. \label{eq:g2}
\end{eqnarray}
Equations (\ref{eq:g0} and \ref{eq:g2}) are accurate to within 1\% and 10\% for $0<s<\infty$, respectively.

The gas phase abundance is defined by $A_i\equiv n_i/n_\mathrm{H}$, where $n_i$ is the number density of species $i$. The abundance is taken from the standard solar elemental abundances \citep{Grevesse98}. In this study, we consider hydrogen, helium, carbon, nitrogen, and oxygen atoms, and their abundances are $A_\mathrm{He}=8.51\times10^{-2}$, $A_\mathrm{C}=3.31\times10^{-4}$, $A_\mathrm{N}=8.32\times10^{-5}$, and $A_\mathrm{O}=6.76\times10^{-4}$. For the sake of simplicity, every ion is assumed to carry $+e$ charge, then $z_i=1$. The electron abundance is $A_\mathrm{e}=1.09$.

Recently, \cite{H17} pointed out if drift velocity exceeds a few percent of the speed of light, Equation (\ref{eq:gasdrag}) breaks down 
because the penetration length of an incident atom becomes comparable to or larger than the dust grain size. Although typical drift speed treated in this study is marginally lower than this threshold, we still use Equation (\ref{eq:gasdrag}) for the sake of simplicity. 

\subsection{Grain Charging} \label{sec:charge}
In order to compute the Coulomb drag force, we need to know grain charge. We compute grain charge $\zd$ by solving the rate equation \citep{WDB06}:
\begin{equation}
\frac{d\zd}{dt}=\jpe-\je+\jsec+\jion, \label{eq:charge}
\end{equation}
where $\jpe$ is the photoelectric emission rate, $\je$ and $\jion$ are the collisional charging rate of electrons and ions, respectively, and $\jsec$ is the secondary gas emission rate (see Appendix \ref{sec:jrate} for more detail).
Photoelectric emission, ion collisions, and secondary gas emission make grains being positively charged, while electron collisions make them negatively charged. 
Typical collision timescale of electrons for a neutral grain is $\je^{-1}\sim0.03$ s, where we have used $\tg=10^5$ K, $a=0.1~\mu$m, $n_e=10^3~\mathrm{cm}^{-3}$, and $s_e=0.5$ (see Equations \ref{eq:je} and \ref{eq:jtil1}). Since charging timescale is much shorter than dynamical timescale, we can assume the steady state for grain charge. In addition, we also ignore charge distribution because grains will be highly charged in the AGN environments \citep{WDB06}. Thus, in this paper, we solve $\jpe-\je+\jsec+\jion=0$ to find $\zd$. 
Solutions of Equation (\ref{eq:charge}) are presented in the companion paper \citep{RT20a} (see also \citet{WDB06}).

\subsection{Sputtering} \label{sec:spt}
\begin{figure*}[t]
\begin{center}
\includegraphics[height=6.0cm,keepaspectratio]{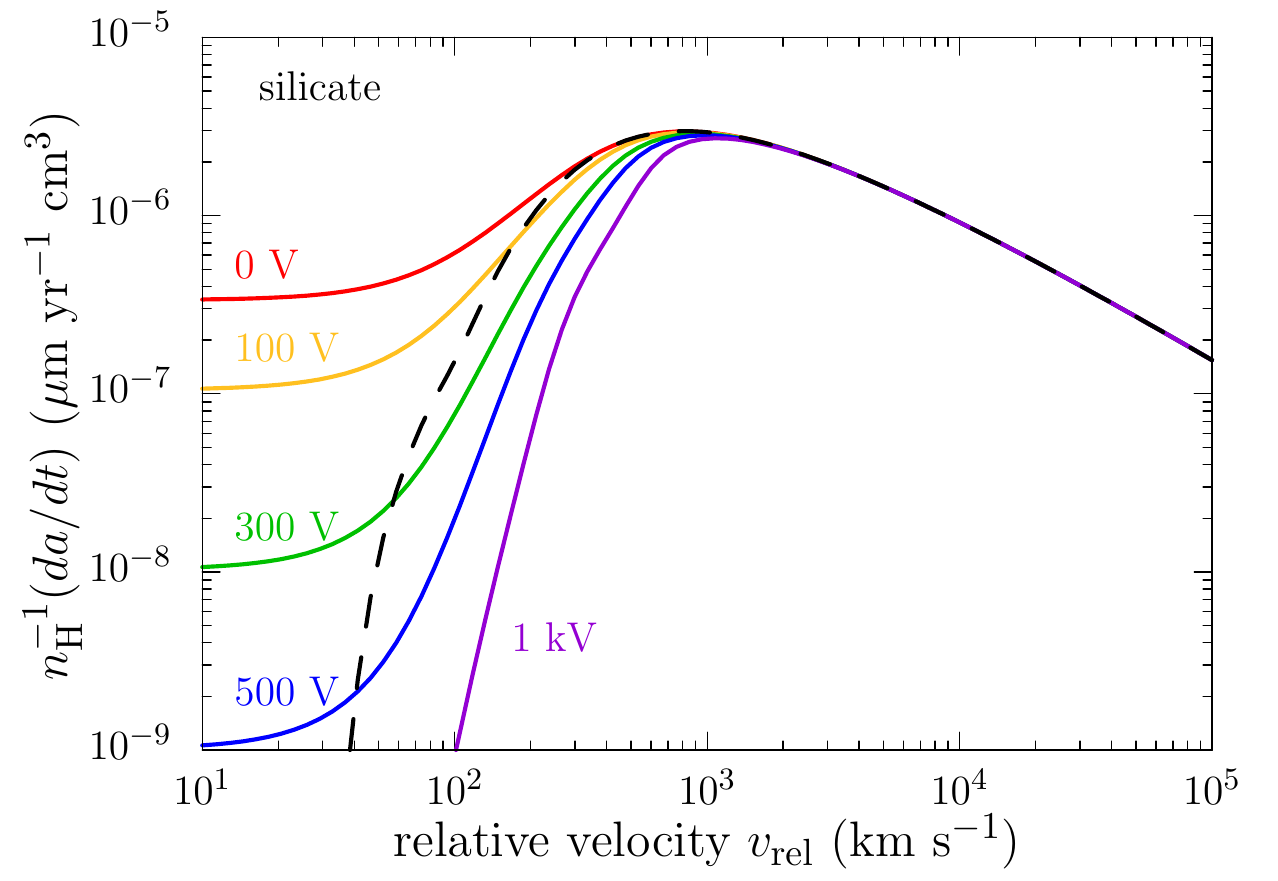}
\includegraphics[height=6.0cm,keepaspectratio]{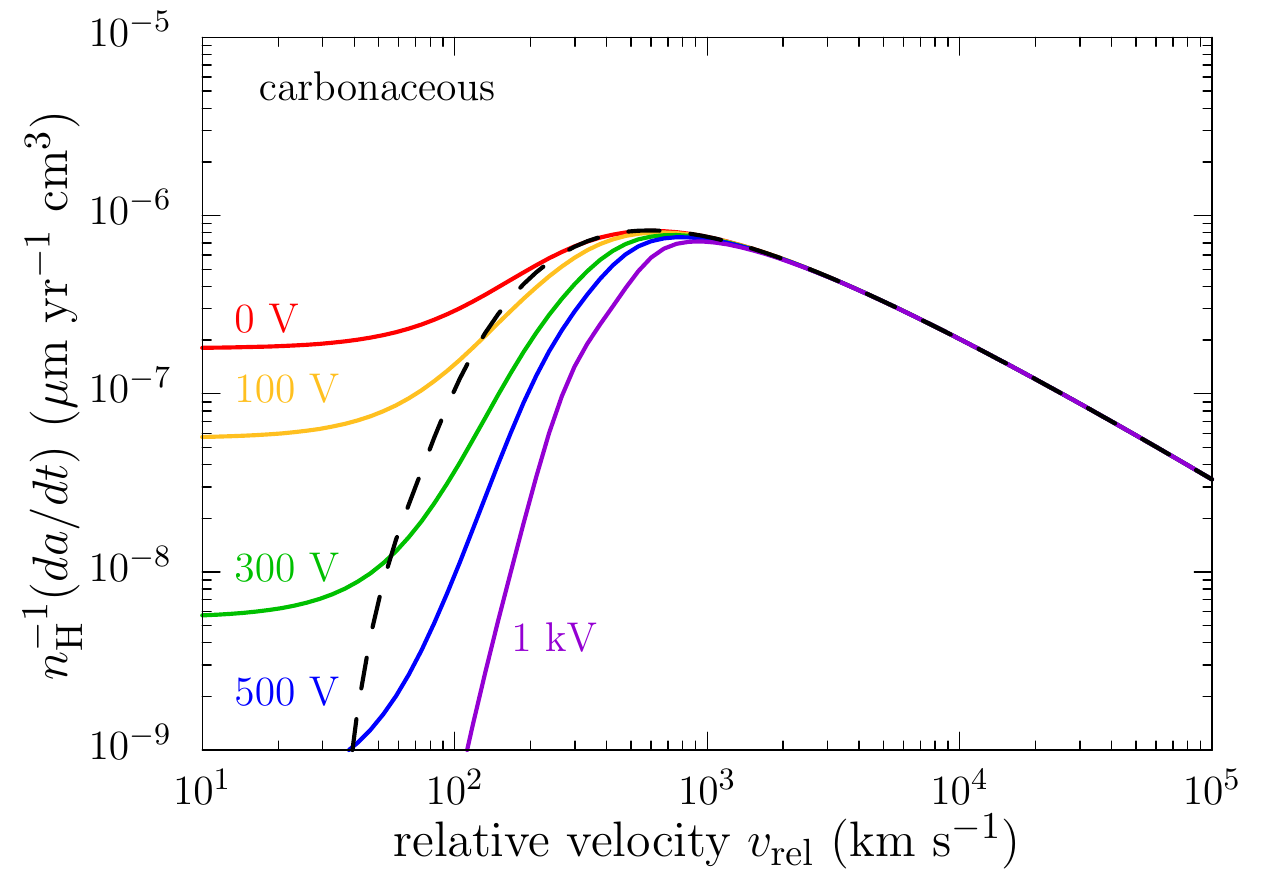}
\caption{Sputtering rate with $\tg=10^6$ K. Left and right panels show the sputtering rate of silicate and carbonaceous grains, respectively. Solid lines are the solution to Equation (\ref{eq:sprate}), where different colors correspond to different grain charge. Dashed lines shows the kinetic sputtering rate ($s\gg1$, see also Equation \ref{eq:spkin}).}
\label{fig:sprate}
\end{center}
\end{figure*}

Dust grains at the vicinity of AGNs may be sputtered by gas either thermally or kinetically (non-thermally). The sputtering rate can be expressed as \citep[e.g.,][]{D79, D92, N06}
\begin{eqnarray}
&&\frac{1}{n_\mathrm{H}}\frac{da}{dt}=-\frac{m_\mathrm{sp}}{2\rho_s}\sum_i \left(\frac{8\kb\tg}{\pi m_i}\right)^{\frac{1}{2}} A_i \frac{e^{-s_i^2}}{2s_i} \nonumber\\
&&\times\int_{\epsilon_\mathrm{min}}^\infty d\epsilon_i \left(1-\frac{z_i\phi}{\epsilon_i}\right)\sqrt{\epsilon_i}e^{-\epsilon_i} \sinh(2\sqrt{\epsilon_i}s_i)Y^i_\mathrm{sp}[(\epsilon_i-z_i\phi)kT], \label{eq:sprate}\nonumber\\
\end{eqnarray}
where $\epsilon_i=E_i/\kb\tg$ is normalized kinetic energy of an impacting ion, $Y^i_\mathrm{sp}$ is the normal-incidence sputtering yield, and $m_\mathrm{sp}$ is the average mass of the sputtered atom. The sputtering yield is calculated based on the model described in \citet{N06} (see also Appendix \ref{sec:ysp}).

Figure \ref{fig:sprate} shows the sputtering rate as a function of the relative velocity. Sputtering can be classified into two origins: thermal sputtering and kinetic (non-thermal) sputtering \citep[e.g.,][]{D92, N06}. The former one is sputtering due to thermal motion of gas particles, whereas the latter is driven by the relative velocity between gas and dust. 
In Figure \ref{fig:sprate}, the sputtering rate at the low velocity domain ($\vrel\ll10^2$ km s$^{-1}$) corresponds to the rate of thermal sputtering.

It can be seen from Figure \ref{fig:sprate} that with increasing grain potential, the thermal sputtering rate decreases. This is because collisions of thermal ions are suppressed due to the Coulomb repulsion force between a grain and the ions. 
In Equation (\ref{eq:sprate}), a collision between a positively charged grain and a hydrogen ion with kinetic energy of about thermal energy ($\epsilon_i\simeq1$) occurs when $z_i\phi/\epsilon_i=eU/\kb\tg\lesssim1$. Thus, thermal sputtering occurs when 
\begin{equation}
    \tg\gtrsim3.5\times10^6~\mathrm{K}\left(\frac{U}{300~\mathrm{V}}\right).
\end{equation}
Since dust grains exposed to the radiation from central engine usually become charged with a few hundreds of volts \citep{WDB06, RT20a}, thermal sputtering is not likely to occur unless the gas temperature is extremely high.

Similarly, kinetic sputtering due to hydrogen-ion collisions occurs when $z_i\phi/\epsilon_i=2eU/m_\mathrm{H}\vrel^2\lesssim1$, where we have used $\epsilon_i=m_\mathrm{H}\vrel^2/2\kb\tg$. As a result, we obtain
\begin{equation}
    \vrel\gtrsim2.4\times10^2~\mathrm{km~s}^{-1}\left(\frac{U}{300~\mathrm{V}}\right)^{1/2}. \label{eq:vkin}
\end{equation}
As we will show, typical relative velocity, which is set by the terminal velocity (see Equation (\ref{eq:vterm})), can be much higher than this velocity when the Coulomb drag force is weaker than the radiation pressure force. Hence, we can expect kinetic sputtering at the vicinity of AGNs.

When kinetic sputtering is dominant, Equation (\ref{eq:sprate}) can be reduced to more simple form.
In the limit of high relative velocity ($s_i\to\infty$), we obtain \citep{D92, N06},
\begin{equation}
\frac{1}{\nh}\frac{da}{dt}=-\frac{m_\mathrm{sp}}{2\rho_s}\vrel\sum_i A_iY^0_i\left(\frac{m_i\vrel^2}{2}\right), \label{eq:spkin}
\end{equation}
where we have ignored grain charge.

\section{Results} \label{sec:results}
In this section, we investigate drift-induced sputtering at the pc-scale regions of AGN. In Section \ref{sec:drag}, we study the condition for the hypersonic drift, which is necessary to cause drift-induced sputtering. In Section \ref{sec:agnrad}, we solve grain dynamics and destruction to show how drift-induced sputtering occurs around AGNs. Section \ref{sec:torusrad} investigates the effect of radiation from the dusty disk on trajectory of a grain and discusses how grains are blown out to the polar region.

\subsection{Grain Drift Velocity: Subsonic or Hypersonic?} \label{sec:drag}

\begin{figure*}[t]
\begin{center}
\includegraphics[height=6.0cm,keepaspectratio]{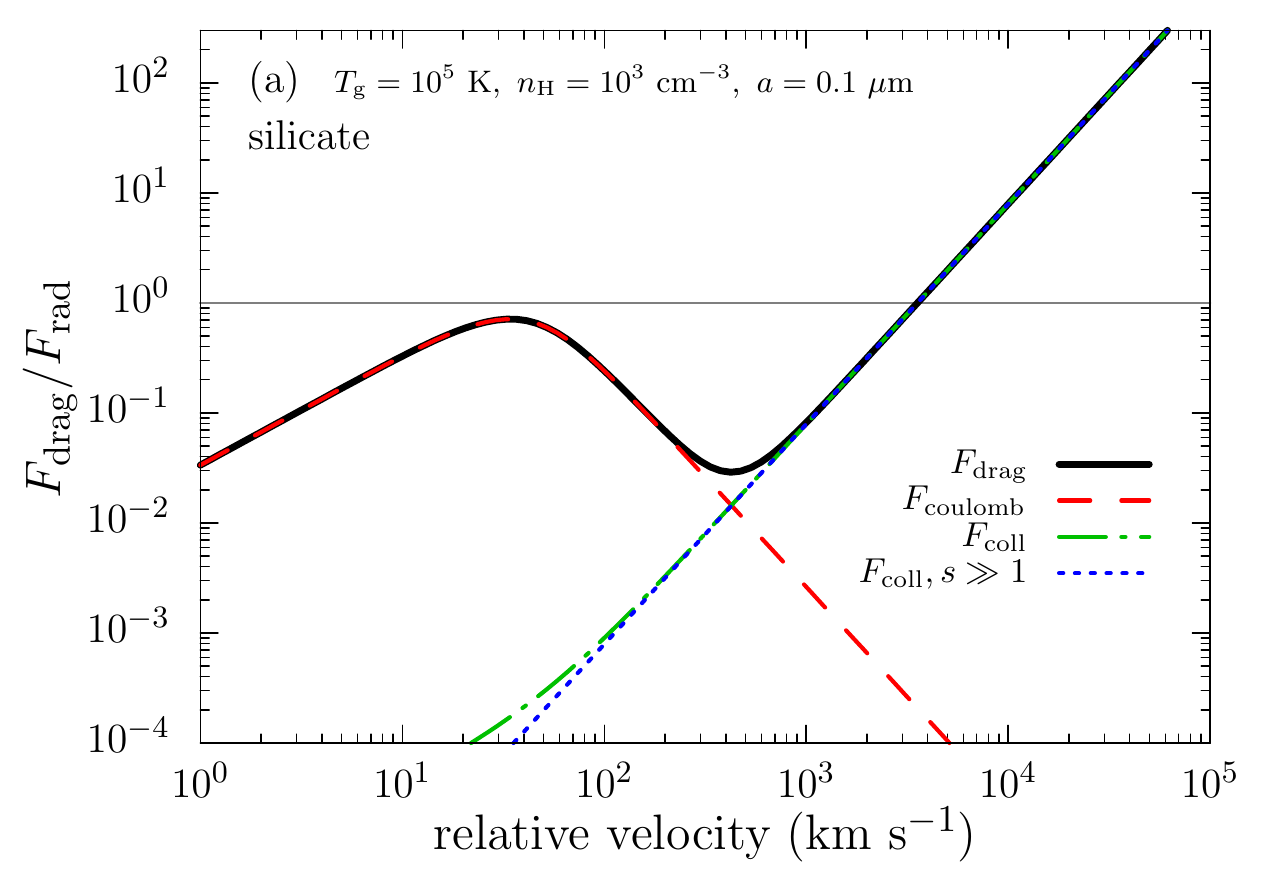}
\includegraphics[height=6.0cm,keepaspectratio]{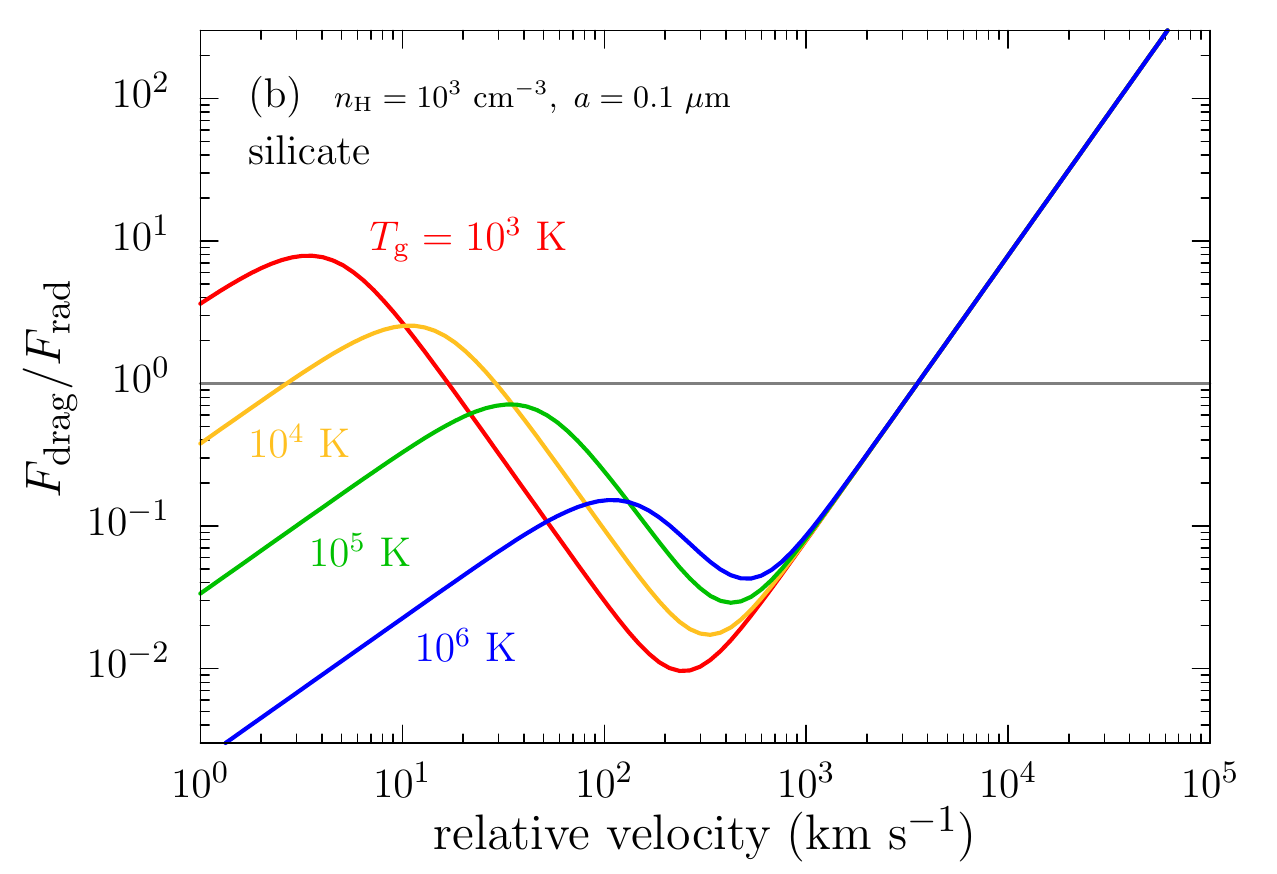}
\includegraphics[height=6.0cm,keepaspectratio]{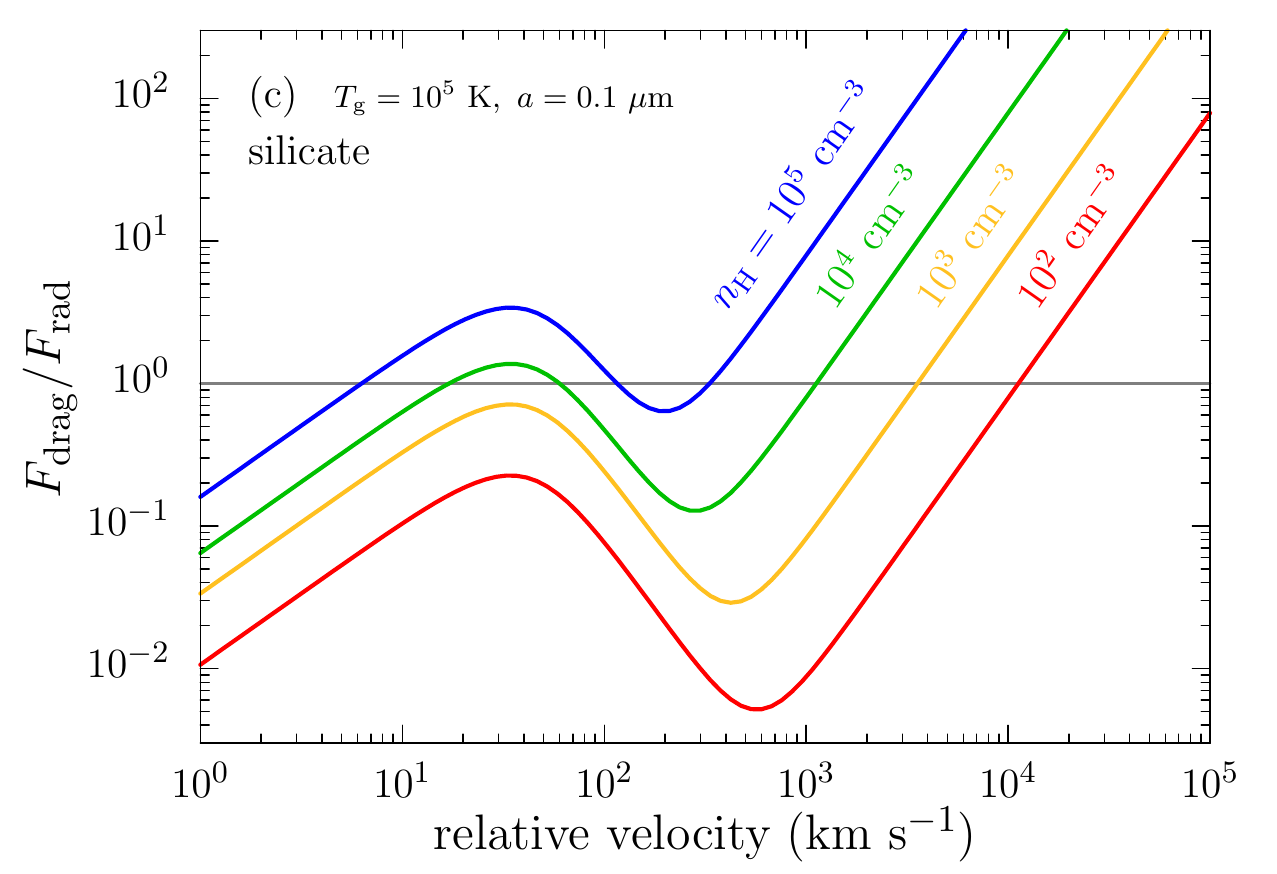}
\includegraphics[height=6.0cm,keepaspectratio]{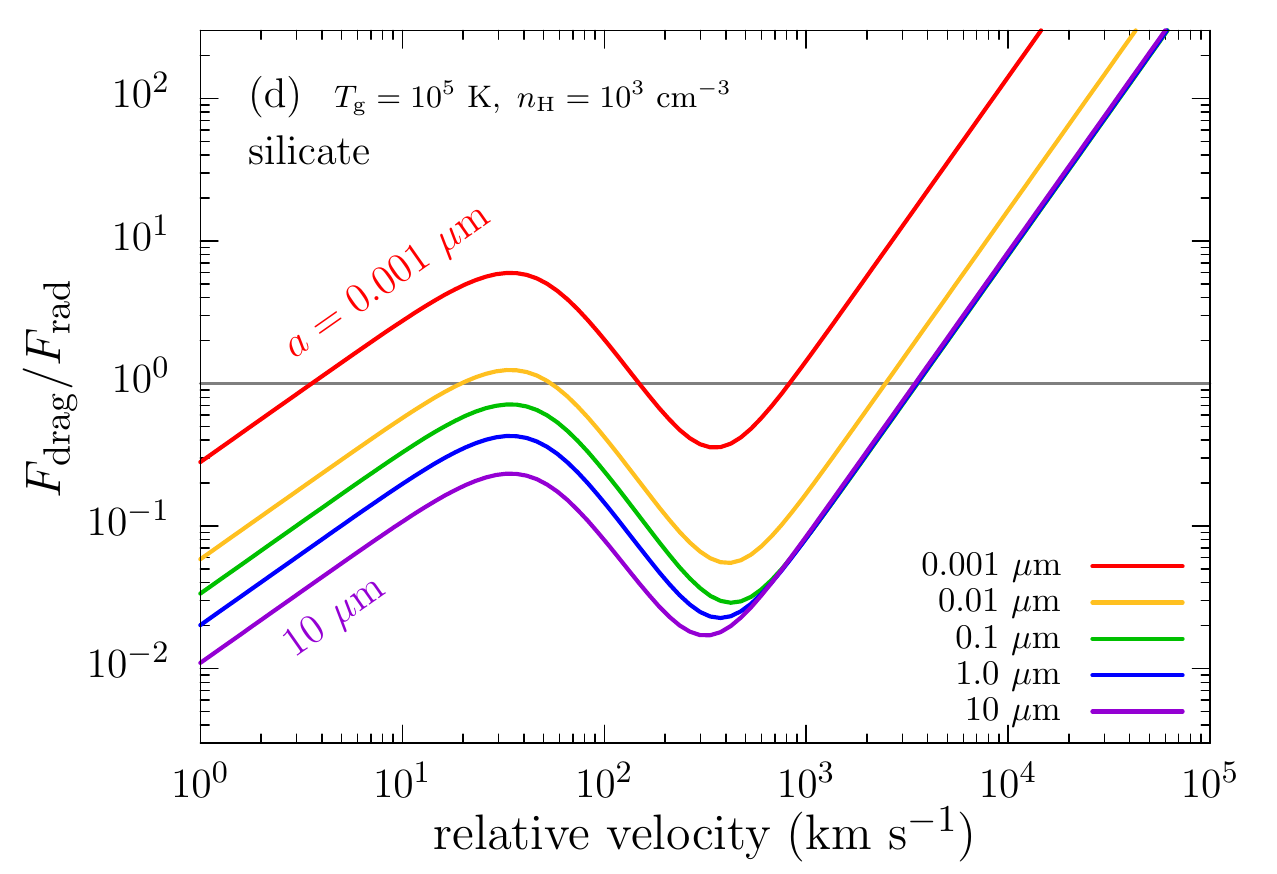}
\includegraphics[height=6.0cm,keepaspectratio]{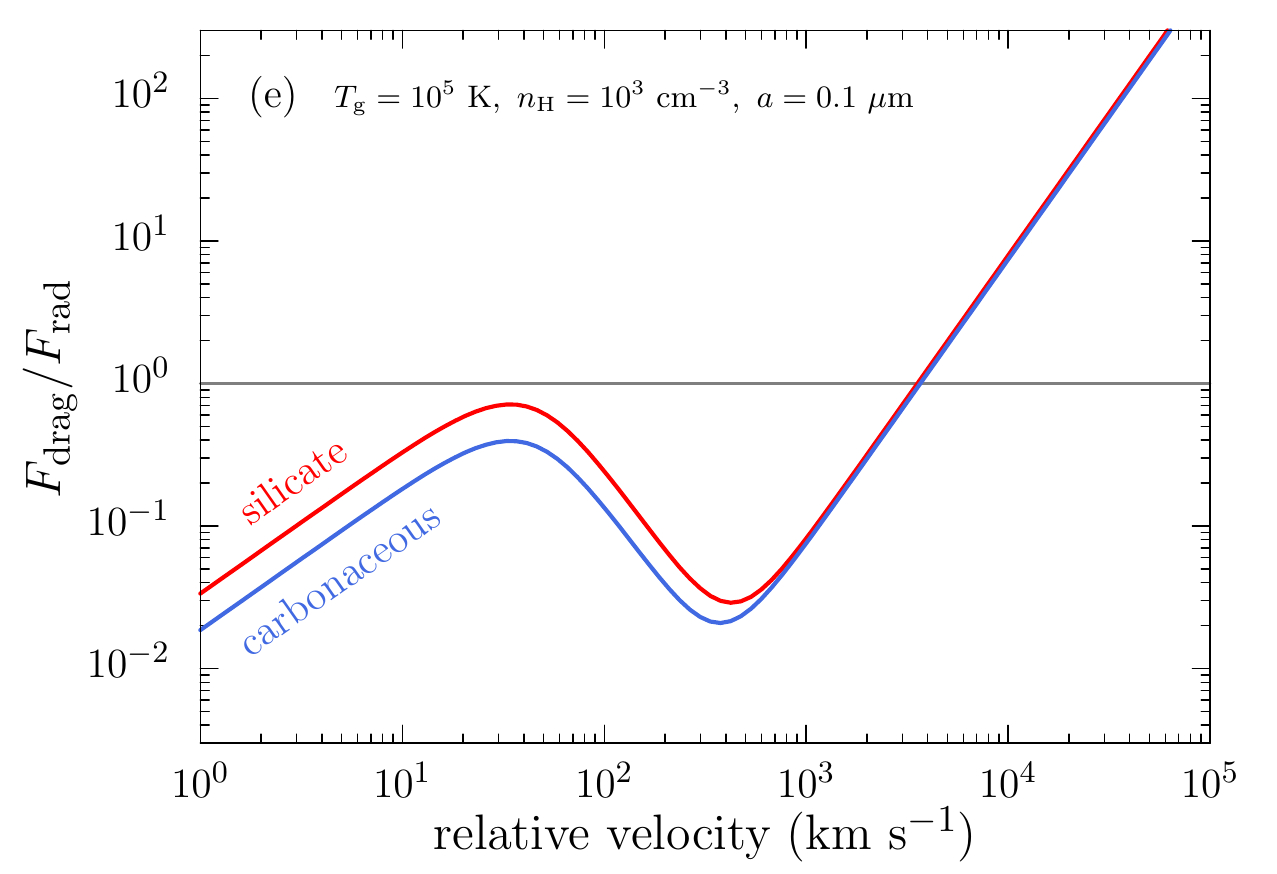}
\includegraphics[height=6.0cm,keepaspectratio]{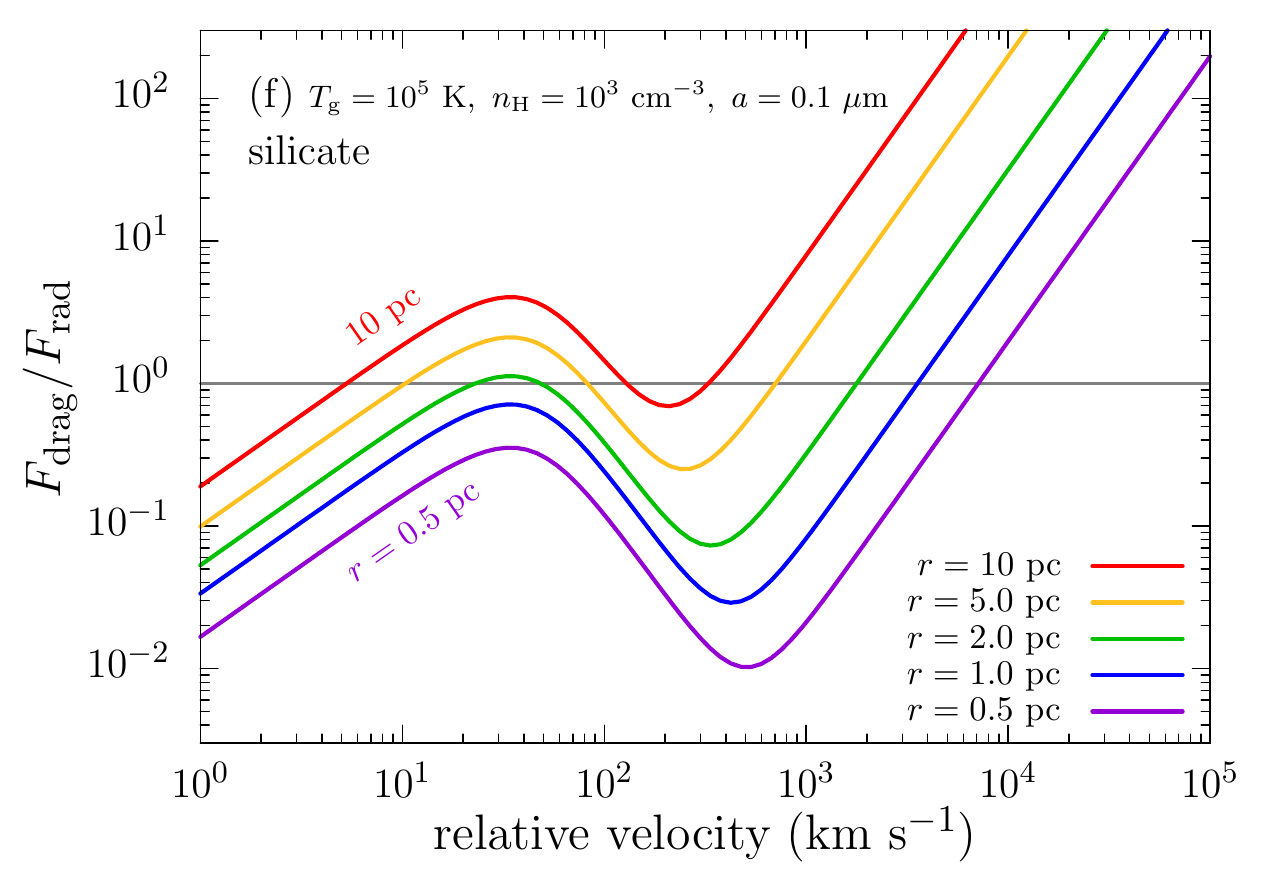}
\caption{Radiation pressure force of AGN radiation vs. gas drag force for various parameters. (a) Comparison between collisional and Coulomb drag forces (b) Gas temperature dependence, (c) gas density dependence, (d) grain radius dependence, (d) grain composition dependence, and (e) initial position dependence.}
\label{fig:fdrag}
\end{center}
\end{figure*}

The terminal velocity of grains will be set by the balance between radiation pressure and gas drag. 
Figure \ref{fig:fdrag} shows the gas drag force relative to the radiation pressure force as a function of the relative velocity for a wide range of parameters. As shown in Figure \ref{fig:fdrag}(a), the gas drag force consists of two contributions: Coulomb drag and collisional drag (see also Section \ref{sec:fdrag}). At the low velocity regime, the Coulomb drag force dominates, while at the high velocity regime, the collisional drag force dominates. 

First of all, we study how the Coulomb drag force varies with parameters.
The Coulomb drag force becomes maximum when the relative velocity approximately coincides with the thermal velocity of gas.
Since the thermal velocity is proportional to $\sqrt{\tg}$, the peak velocity at which the Coulomb drag force is maximized shifts toward larger velocities for higher gas temperature (Figure \ref{fig:fdrag}b). In addition, the Coulomb drag force is weaker for higher gas temperature.
With increasing the gas temperature, a grain will be more positively charged, which makes Coulomb drag strong; however, in the same time, kinetic energy of gas particles is also increased, which makes Coulomb scattering difficult. Since the latter effect dominates the former one, the Coulomb drag force becomes weak for higher gas temperature. 

The Coulomb drag force also increases as the gas density due to more frequent collisions (Figure \ref{fig:fdrag}c), although the Coulomb drag force is not simply proportional to the gas density. This is because higher gas density provides more electrons in the gas phase, which results in reducing the grain charge. 

The Coulomb drag force is weaker for smaller particles because of larger normalized grain potential $\phi=e^2\zd/\kb\tg{a}$ (Figure \ref{fig:fdrag}d). 
$\phi$ usually increases 
with decreasing grain radius as long as $a\gtrsim0.1~\mu$m as suggested in Figure \ref{fig:fdrag}(d) \citep[see also][]{WDB06}.  
In Figure \ref{fig:fdrag}(d), the $\fdrag/\frad$-values suddenly increases for grains smaller than $0.01~\mu$m. 
This is mainly due to the reduction of the radiation pressure force with respect to the one expected from Equation (\ref{eq:beta}), which happens when $a\lesssim0.1~\mu$m (see Figure \ref{fig:beta}). Silicate grains are slightly more charged than carbonaceous grains, and then, the Coulomb drag force is slightly larger for silicate grains (Figure \ref{fig:fdrag}e). The ratio also depends on the initial distance, $\rinit$. As $\rinit$ increases, the ratio increases due to reduction of radiation pressure forces.

Next, we study the collisional drag force.
In the limit of $s\gg1$, the drag force (Equation \ref{eq:drag}) due to direct collision can be reduced to 
\begin{equation}
F_\mathrm{coll,s\gg1}\simeq \sigd n_\mathrm{H}v_\mathrm{rel}^2\sum_i m_iA_i.
\end{equation}
By using Equation (\ref{eq:frad}), we find
\begin{equation}
\frac{F_\mathrm{coll,s\gg1}}{F_\mathrm{rad}}\simeq \frac{cn_\mathrm{H}v_\mathrm{rel}^2\sum_i m_iA_i}{F_\mathrm{AGN}\langle \qpr \rangle_\mathrm{AGN}}, \label{eq:fcoll}
\end{equation}
where $F_\mathrm{AGN}=L_\mathrm{AGN}/4\pi\rp^2$ is the AGN radiation flux. Therefore, as long as $\langle \qpr \rangle_\mathrm{AGN}\approx1$, which becomes good approximation for grains larger than $\approx0.1~\mu$m (see Figure \ref{fig:beta}), Equation (\ref{eq:fcoll}) does not depend on any grain properties, such as size and composition, as well as the gas temperature, as shown in Figure \ref{fig:fdrag}(b,d,e).
Note that in Figure \ref{fig:fdrag}(e), a very subtle difference in 
$F_\mathrm{coll,s\gg1}/F_\mathrm{rad}$ between silicate and carbonaceous grains can be seen. This is due to the slight difference of $\langle \qpr \rangle_\mathrm{AGN}$ arising from their difference of optical constants.

%
%
Finally, we discuss the terminal velocity of a grain.
As mentioned, Coulomb drag is maximized when the relative velocity is similar to the thermal velocity of gas. Therefore, if radiation pressure balances with Coulomb drag, the terminal velocity becomes subsonic. On the other hand, if radiation pressure is stronger than Coulomb drag, and then it finally balances with collisional drag, the terminal velocity will be hypersnoic. 

In the case of the hypersonic drift, the terminal velocity can be found from Equation (\ref{eq:fcoll}) by setting the left-hand side as unity, and it becomes
\begin{eqnarray}
\vterm&=&\left(\frac{F_\mathrm{AGN}\langle \qpr \rangle_\mathrm{AGN}}{cn_\mathrm{H}\sum_i m_iA_i}\right)^{1/2}, \label{eq:vterm}\\
&\simeq&3.5\times10^3~\mathrm{km~s}^{-1}\langle \qpr \rangle_\mathrm{AGN}
\left(\frac{\nh}{10^3~\unitnum}\right)^{-\frac{1}{2}} \nonumber \\
&&\times\left(\frac{L_\mathrm{AGN}}{10^{45}~\mathrm{erg~s}^{-1}}\right)^{1/2}
\left(\frac{\rp}{1~\mathrm{pc}}\right)^{-1}.
\end{eqnarray}
For $a\gtrsim0.1~\mu$m, we have $\langle \qpr \rangle_\mathrm{AGN}\approx1$, and hence, the terminal velocity does not depends on grain radius.
Since this velocity is higher than the threshold velocity of kinetic sputtering (Equation \ref{eq:vkin}), kinetic sputtering is found to be important.

\begin{figure}[t]
\begin{center}
\includegraphics[height=6.5cm,keepaspectratio]{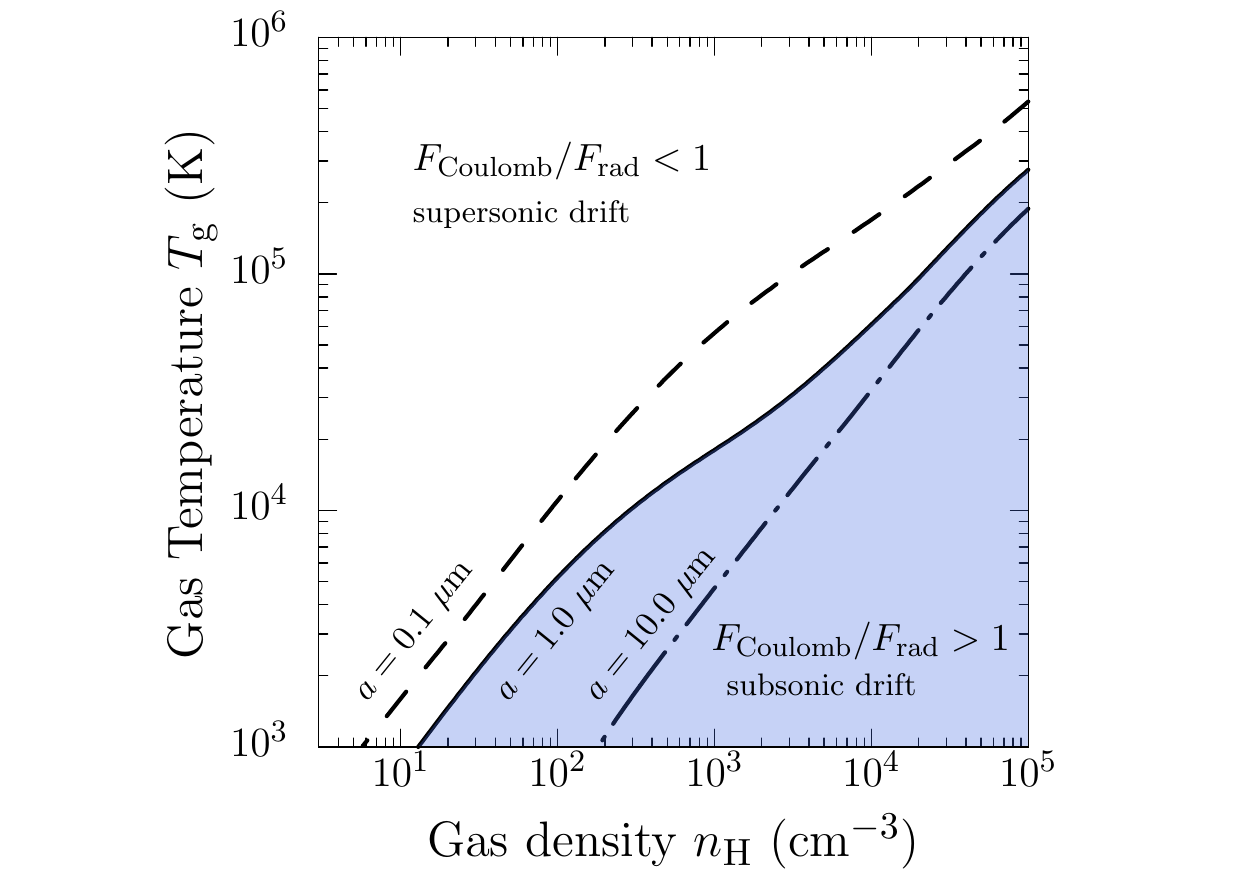}
\includegraphics[height=6.5cm,keepaspectratio]{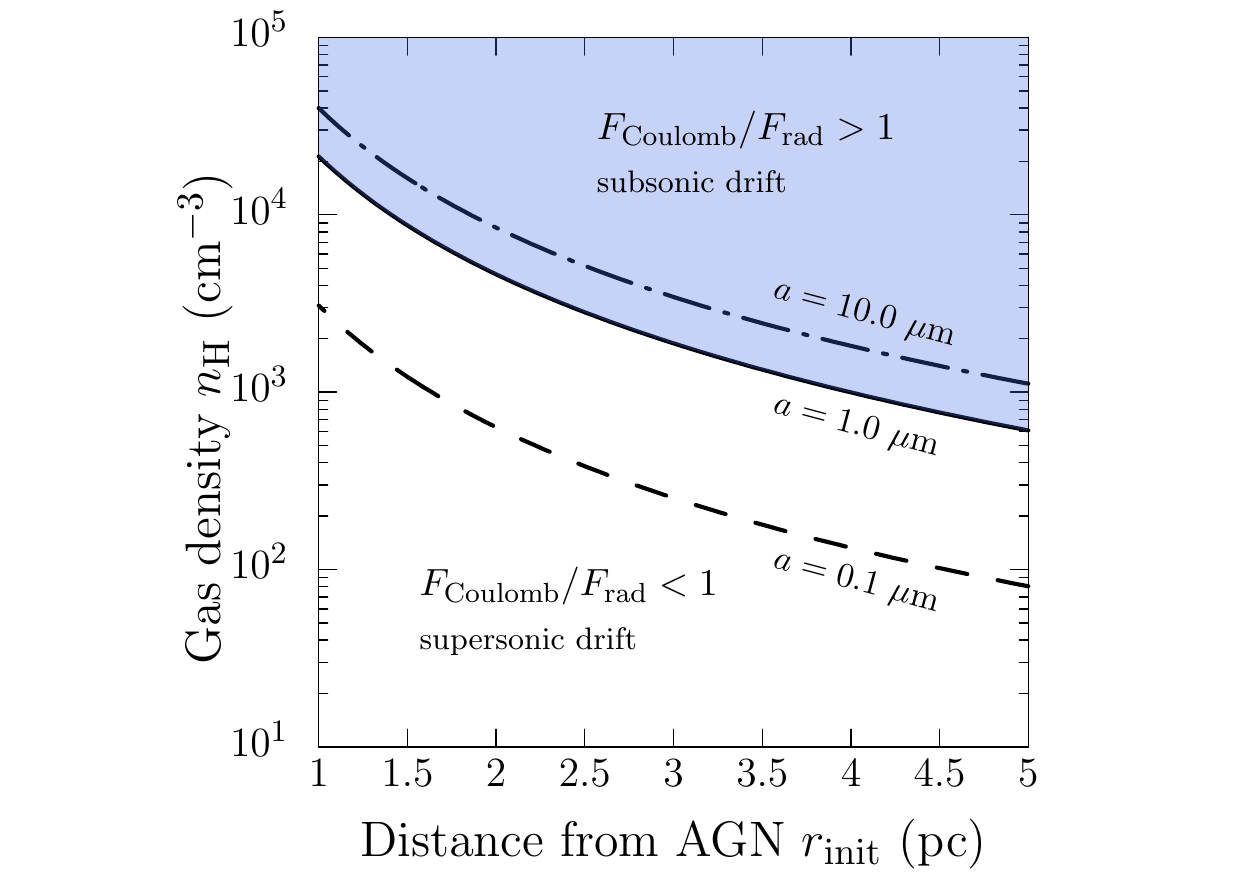}
\caption{The boundary between the hyper and sub-sonic drift. Solid lines are the boundaries for $1~\mu$m-sized grains, whereas dashed and dot-dashed lines are those of $0.1~\mu$m and $10~\mu$m, respectively. Blue-shaded region shows the parameter space where the subsonic drift occurs. Dust composition is assumed to be silicate. (Top) Gas density and temperature dependence. $\rinit=1$ pc is assumed. (Bottom) Gas density and initial located dependence. $\tg=10^5$ K is assumed.}
\label{fig:cond}
\end{center}
\end{figure}

Figure \ref{fig:cond} shows the parameter range where the sub/hyper-sonic drift occurs. As mentioned, Coulomb drag becomes weaker for higher gas temperature and/or lower density. The boundary of the sub/hyper-sonic drift depends on the grain radius. 
At the pc-scale polar regions of AGN, the gas temperature can be heated higher than $10^4$ K, and the gas density are less than $10^3~\unitnum$ \citep[e.g.,][]{W16}. For example, for $\tg=10^4$ K, $\nh=10^2~\unitnum$ and $\rinit=1~$ pc, grains larger than 0.1 $\mu$m will be subjected to the hypersonic drift, whereas grains smaller than $0.1~\mu$m results in the subsonic drift. Although very small grains can avoid destruction by sputtering, they might be disrupted by Coulomb explosion \citep{RT20a}.
In addition, the boundary also depends on the initial distance of a grain from the AGN. 
As shown in Figure \ref{fig:cond} (bottom), for $\nh\lesssim80~\unitnum$, grains larger than 0.1$~\mu$m initially located within 5 pc from AGN can be accelerated to the hypersonic velocity.

Although we have treated gas temperature and gas density as individual parameters, those could be related each other. It is important to study how they are related each other in the AGN environment \citep[e.g.,][]{s14} and how this affects grain dynamics. However, this is beyond the scope of this paper.

\subsection{Drift-Induced Kinetic Sputtering} \label{sec:agnrad}
\begin{figure}[htp]
\begin{center}
\includegraphics[height=6.0cm,keepaspectratio]{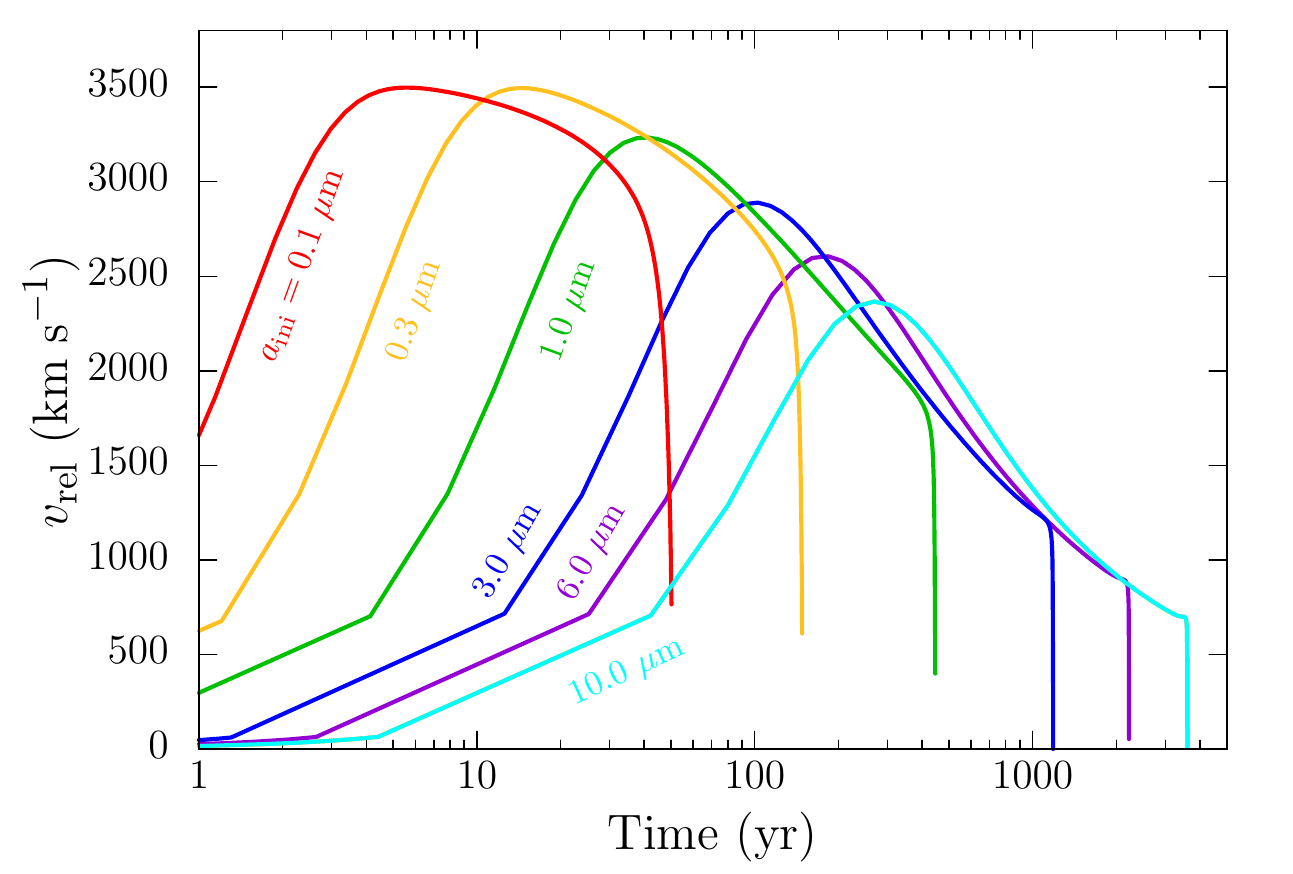}
\caption{Relative velocity vs. time. We set the Eddington ratio of $\lambda_\mathrm{edd}=0.08$ and the gas density of $n_\mathrm{H}=10^3$ cm$^{-3}$. Silicate grains are assumed.}
\label{fig:vt}
\end{center}
\end{figure}

\begin{figure*}[htp]
\begin{center}
\includegraphics[height=12.0cm,keepaspectratio]{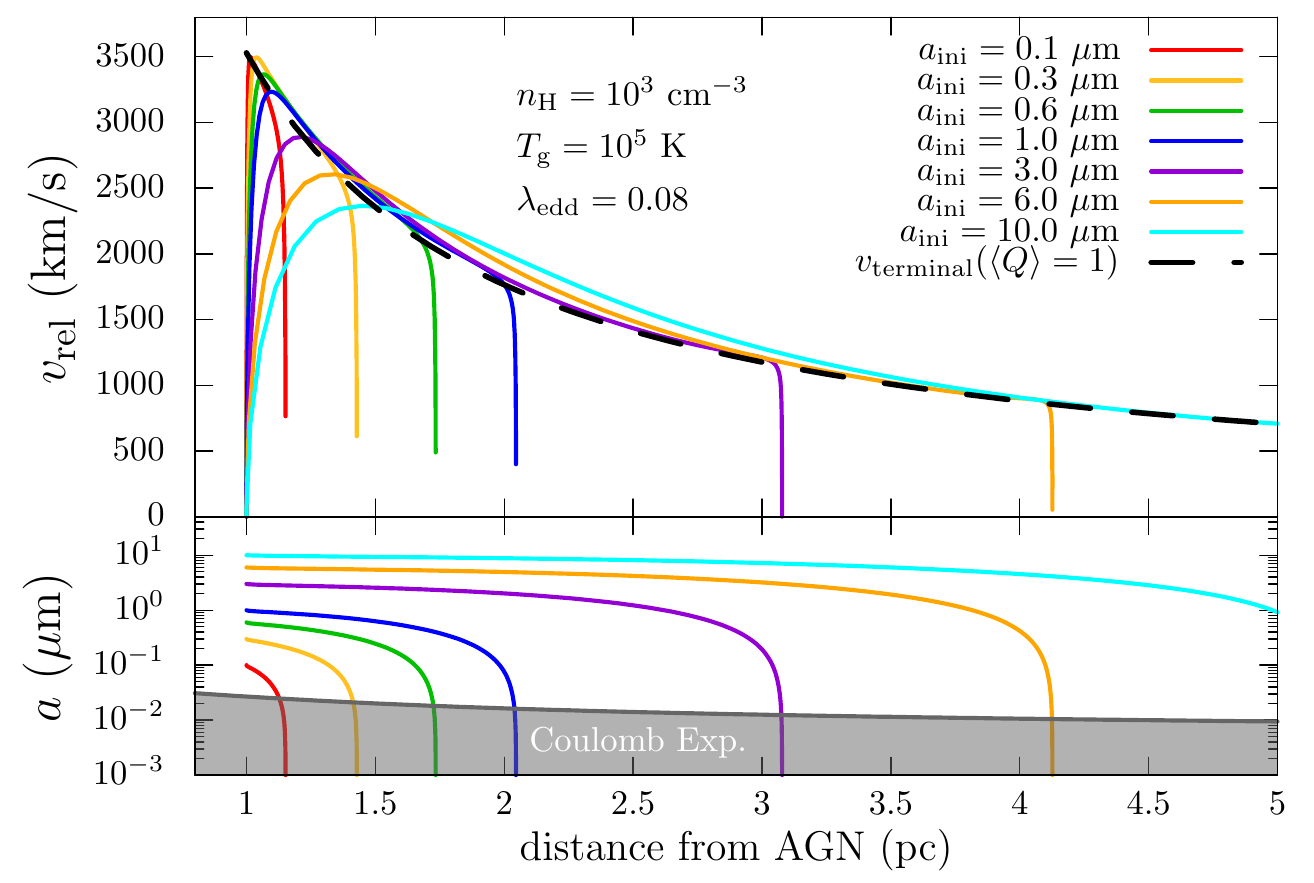}
\caption{Relative velocity (top) and grain radius (bottom) of dust grains with AGN radiation pressure and kinetic sputtering with an Eddington ratio of $\lambda_\mathrm{edd}=0.08$ and a gas density of $n_\mathrm{H}=10^3$ cm$^{-3}$. Dust composition is assumed to be silicate, and its tensile strength is set as $10^{10}\unitb$. The grey shaded region indicates where Coulomb explosion disrupts dust grains \citep[see][for more detail]{RT20a}. Silicate grains are assumed.}
\label{fig:trac1}
\end{center}
\end{figure*}

In order to study dynamics and dust destruction, we solve the equation of motion (Equations \ref{eq:eom1} and \ref{eq:eom2}), grain charge (Equation \ref{eq:charge} with steady state assumption), and sputtering (Equation \ref{eq:sprate}), simultaneously. 
Here, we ignore IR radiation from the dusty disk, and therefore, grain motion becomes one dimensional (radial).
The gas density and temperature are assumed to be $\nh=10^3~\unitnum$ and $\tg=10^5$ K, respectively. Thus, sub-micron-sized or larger grains at $\rinit=1~$pc are subjected to the hypersonic drift (Figure \ref{fig:cond}).

Figure \ref{fig:vt} shows the time evolution of the grain drift velocity. Initially, dust grains are accelerated by the radiation pressure and reach the terminal velocity (see Equation \ref{eq:vterm}). The acceleration time to reach the terminal velocity is typically about
\begin{eqnarray}
\tacc&\equiv&\frac{\md\vterm}{\frad},\\
&\simeq&19~\mathrm{yr}
\langle \qpr \rangle_\mathrm{AGN}^{-\frac{1}{2}}
\left(\frac{a}{1~\mu\mathrm{m}}\right)
\left(\frac{\nh}{10^3~\unitnum}\right)^{-\frac{1}{2}} \nonumber\\
&&
\left(\frac{L_\mathrm{AGN}}{10^{45}~\mathrm{erg~s}^{-1}}\right)^{-1/2}
\left(\frac{\rp}{1~\mathrm{pc}}\right).
\end{eqnarray}
Thus, larger grains require more time to reach the terminal velocity. Once the grains reach the terminal velocity, grain radius gradually decreases as they drift because the relative velocity is high enough to cause sputtering (Figure \ref{fig:trac1}). 
The sputtering timescale is typically about
\begin{eqnarray}
\tspt&\equiv&\left[\frac{1}{a}\left(\frac{da}{dt}\right)\right]^{-1},\\
&\simeq&6.6\times10^2~\mathrm{yr}
\left(\frac{a}{1~\mu\mathrm{m}}\right)
\left(\frac{m_\mathrm{sp}}{20~m_\mathrm{H}}\right)^{-1} \nonumber \\
&&\times
\left(\frac{\nh}{10^3~\unitnum}\right)^{-1}
\left(\frac{\vrel}{10^3~\mathrm{km~s}^{-1}}\right)^{-1}
\left(\frac{Y_\mathrm{sp}}{10^{-2}}\right)^{-1}
\end{eqnarray}
As long as $\langle \qpr \rangle_\mathrm{AGN}\approx1$, the terminal velocity (or the relative velocity) does not depend on grain radius, and then, we have $\tspt\propto a$. 
Therefore, smaller grains are more rapidly eroded compared with larger grains. 
Once the grain radius becomes below about $0.1~\mu$m, the grain drift velocity rapidly decreases as grain radius decreases. This is because the size-dependent $\langle \qpr \rangle_\mathrm{AGN}$ is decreased. Finally, dust grains will be disrupted by Coulomb explosion, where grain charge causes grain fission \citep[see][for more detail]{RT20a}. 

\begin{figure}[htp]
\begin{center}
\includegraphics[height=6.0cm,keepaspectratio]{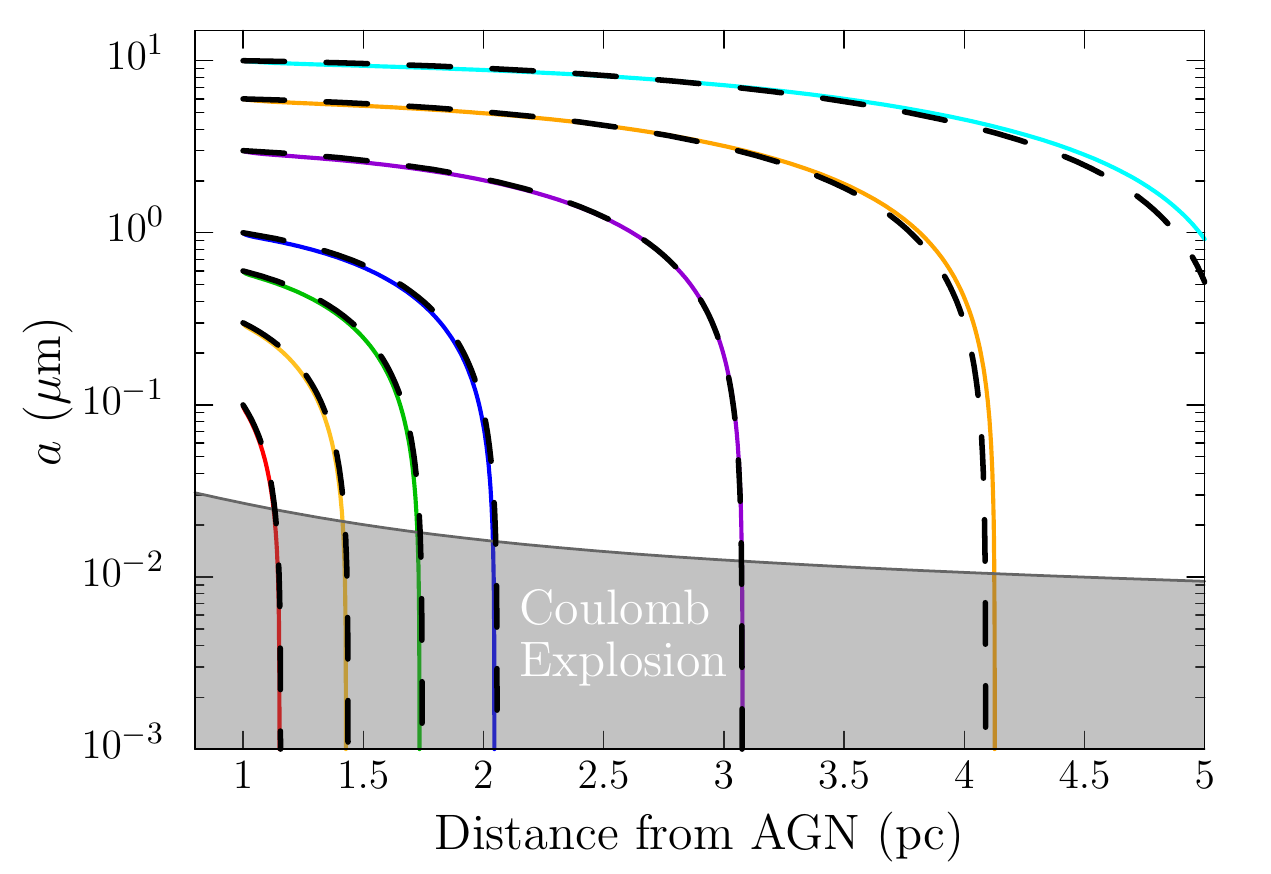}
\includegraphics[height=6.0cm,keepaspectratio]{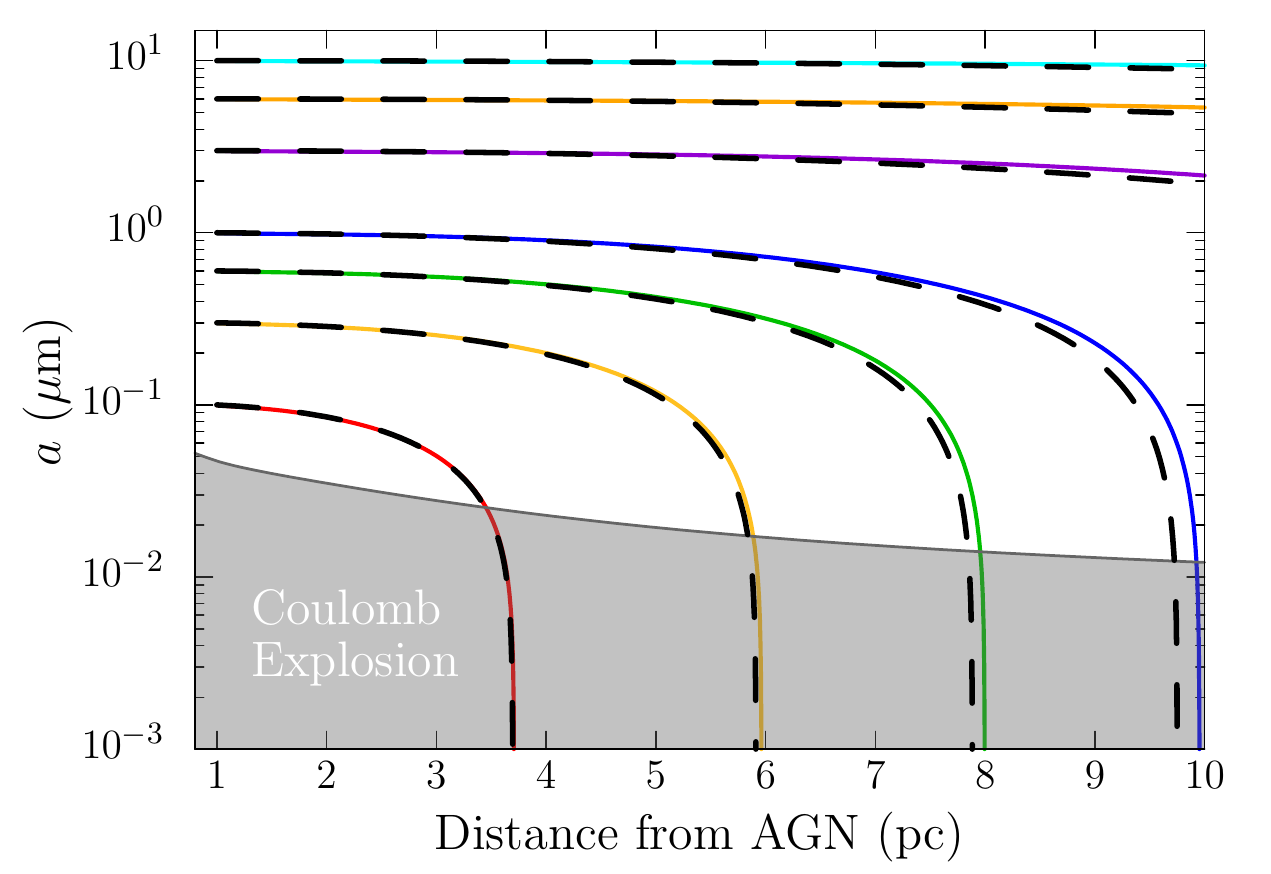}
\caption{Same as Fgure \ref{fig:trac1} (bottom). Dashed lines are obtained by the simplified relation (Equation \ref{eq:sim}). (Top) $\nh=10^3~\unitnum$ and $\tg=10^5$ K. (Bottom) $\nh=10^2~\unitnum$ and $\tg=10^5$ K. Silicate grains are assumed.}
\label{fig:trac2}
\end{center}
\end{figure}

Evolution of grain radius as a function of the distance from AGN can be approximately reproduced by using a simplified equation. 
By using Equations (\ref{eq:eom1} and \ref{eq:spkin}), we can eliminate time derivative, and we obtain a simple equation:
\begin{eqnarray}
\frac{da}{d\rp}&=&-
\frac{m_\mathrm{sp}}{2\rho_s}\nh \sum_i A_iY^0_i\left(\frac{m_i\vterm^2}{2}\right), \label{eq:sim}
\end{eqnarray}
where we have substituted the terminal velocity as the relative velocity and ignored grain charge.
We solve Equation (\ref{eq:sim}), and the results are shown in Figure \ref{fig:trac2} with dashed lines. Equation (\ref{eq:sim}) is able to capture overall properties of drift-induced kinetic sputtering. A distance for which a grain can travel depends on the gas density; a lower gas density increases the travel distance (see Figure \ref{fig:trac2}).

\begin{figure}[htp]
\begin{center}
\includegraphics[height=6.5cm,keepaspectratio]{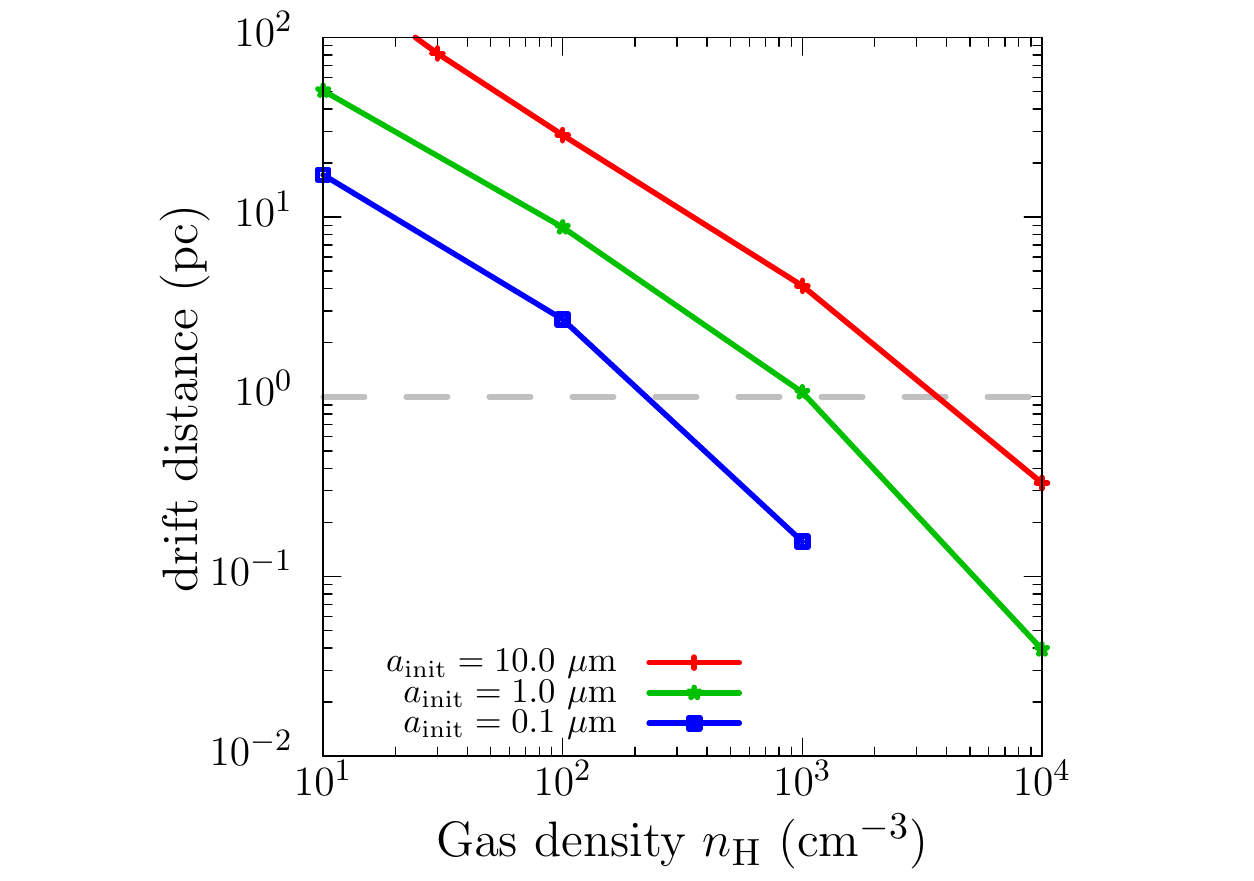}
\caption{Drift distance as a function of gas density for various initial grain radius $a_\mathrm{int}$ obtained from Equation (\ref{eq:sim}). Initial distance and gas temperature are assumed to be 1 pc and $\tg=10^5$ K, respectively. Blue, green, and red lines represent initial grain radii of 0.1, 1.0, and 10.0 $\mu$m, respectively. For the initial grain radius of 0.1 $\mu$m, no point is plotted at gas density of $10^4~\mathrm{cm}^{-3}$ because grain drift is suppressed by Coulomb drag (see Figure \ref{fig:cond}). Silicate grains are assumed. Dashed gray horizontal line represents the drift distance of 1~pc.}
\label{fig:rdead}
\end{center}
\end{figure}

Figure \ref{fig:rdead} shows the drift distance defined by the distance between initial and final distances from AGN, where the final distance is the location when grains radius becomes $10^{-3}~\mu$m (see also the bottom panel of Figure~\ref{fig:trac1}). With increasing gas density, drift distance decreases due to efficient sputtering. 
At $n_\mathrm{H} \lesssim 20$~cm$^{-3}$, both submicron and micron-sized grains drift longer than pc scales. On the other hand, 
At $n_\mathrm{H} \gtrsim 4\times10^3$~cm$^{-3}$,
10 micron-sized grains are disrupted by kinetic sputtering within sub-pc scales. As a result, intermediate density ($n_\mathrm{H}\sim 10^{3}~\mathrm{cm}^{-3}$) is favorable for submicron/micron-sized grains to drift less/more than pc scales as long as gas temperature is higher than $10^5~$K. If gas temperature is lower than that, further lower gas density is necessary to occur drift-induced sputtering because Coulomb drag halts grain drift (see Figure \ref{fig:cond}).

\begin{figure*}[htp]
\begin{center}
\includegraphics[height=8.0cm,keepaspectratio]{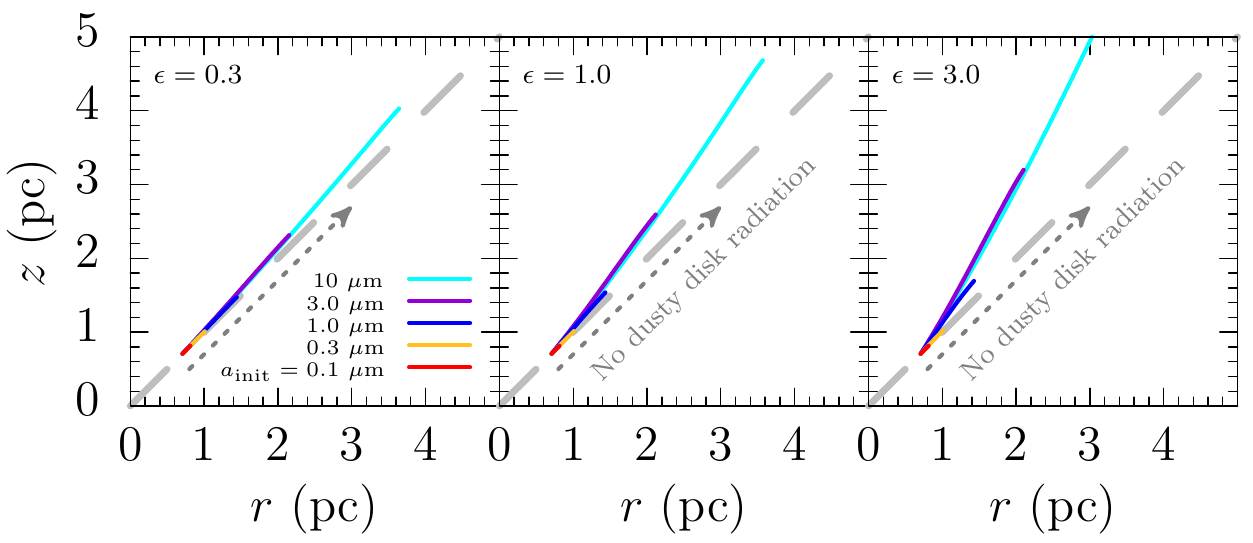}
\caption{Trajectory of dust grains for $\lambda_\mathrm{edd}=0.08$ and $n_\mathrm{H}=10^3$ cm$^{-3}$. From left to right panels, dusty disk luminosity changes with $\epsilon=0.3$, $1.0$, and $3.0$, respectively. The dashed line and arrow represent the blown-out direction without
radiation pressure from the dusty disk. Each solid line corresponds to a different initial grain radius value and terminates once the grain radius becomes less than 0.001 $\mu$m due to sputtering.}
\label{fig:trac3}
\end{center}
\end{figure*}

\subsection{Effect of Dusty Disk Emission on Trajectory} \label{sec:torusrad}
Next, we consider IR radiation from the dusty disk, in addition to AGN radiation pressure, and study how 
dusty disk emission levitates grains. Figure \ref{fig:trac3} presents the trajectories of dust grains with various initial grain radii. Each line terminates once the grain radius becomes less than $10^{-3}~\mu$m.  As shown in Section \ref{sec:agnrad}, larger grains are blown over longer distances than smaller grains. As the torus luminosity ($\epsilon$) increases, dust grains can levitate to higher altitudes. 
Additionally, larger grains can be lifted to higher altitudes, due to the larger radiation pressure cross-section to IR emission from the dusty disk.
For example, $Q_\mathrm{pr}$ of torus IR emission for 3-$\mu$m grains
is larger than that of 0.1-$\mu$m grains. 
In contrast, because both 10- and 3-$\mu$m grains have $Q_\mathrm{pr}\simeq1$, the radiation pressure per unit mass for 10-$\mu$m grains is smaller than that of 3-$\mu$m grains. As a result, the levitation of 10-$\mu$m grains is less efficient than 3-$\mu$m grains.

\begin{figure}[t]
\begin{center}
\includegraphics[height=7.0cm,keepaspectratio]{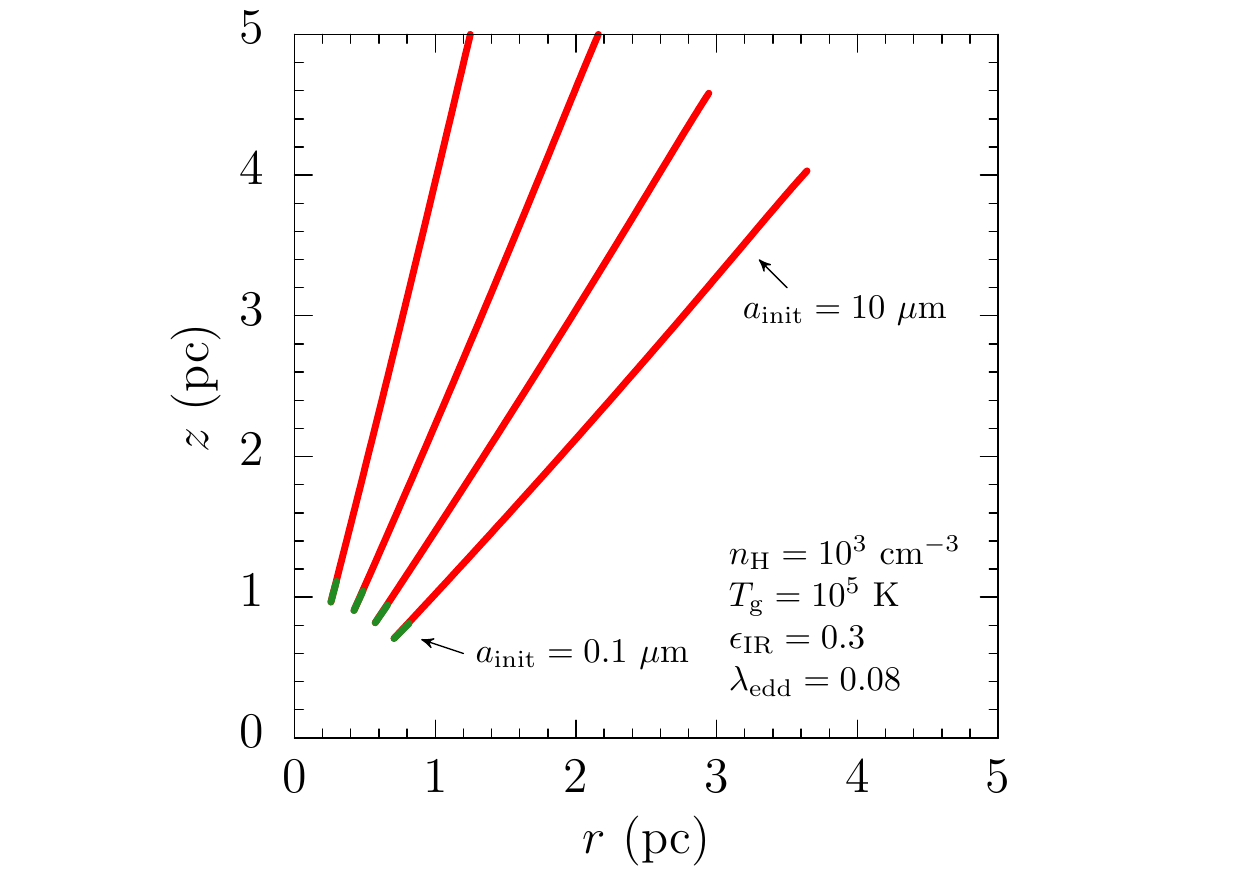}
\caption{Trajectory of $0.1\ \mu$m (green) and 10 $\mu$m (red) grains, initially located in the inner polar region of the AGN.}
\label{fig:trac4}
\end{center}
\end{figure}

We also found that the levitation due to torus emission is a minor effect, as AGN luminosity is usually higher than
that of a dusty disk 
\citep[e.g.,][]{elv94,ric06,bro06}
with $\epsilon=0.1-0.3$, indicating that 
the IR emission from the dusty disk may not be a crucial carrier of dust grains toward polar regions. If dust grains are distributed in the inner polar region, one can readily imagine that they will be blown-out to the polar region. Figure \ref{fig:trac4} presents trajectories of grains initially located in the inner polar regions. Small grains become sputtered after traveling small distances, whereas larger grains can be blown over larger distances. Hence, if the polar dust grains originate from the inner polar region, the drift-induced sputtering scenario predicts a lack of small grains, leading to the suppression of 10-$\mu$m silicate in MIR emission spectra.

\section{Discussion} \label{sec:discussion}
\subsection{Grain Growth in AGN environments}
In Section \ref{sec:results}, we implicitly assumed the presence of micron-sized grains in the inner region of a dusty disk. Here, we discuss the possibility of grain growth 
in the AGN environments.

\subsubsection{Growth by Coagulation in the Dusty Disks}
There are two types of grain growth exist: condensation and coagulation.

Condensation (and evaporation) rate of a spherical grain can be estimated by the Hertz-Knudsen equation: 
\begin{equation}
\frac{da}{dt}=s_\mathrm{a}V_0 \frac{(p-p_\mathrm{sat})}{\sqrt{2\pi m_\mathrm{atom} \kb\tg}}, \label{eq:hk}
\end{equation}
where $p$ and $p_\mathrm{sat}$ are the vapor pressure and saturated vapor pressure, respectively, $s_\mathrm{a}$ is the sticking probability, $m_\mathrm{atom}$ is the mass of impinging atoms, and $V_0=m_\mathrm{atom}/\rho_s$ \citep[e.g.,][]{Henning10, N13}.
In Equation (\ref{eq:hk}), we have assumed dust temperature equals to gas temperature. If the supersaturation ratio, $\mathcal{S}=p/p_\mathrm{sat}$, exceeds unity, condensation happens. The condensation timescale is given by
\begin{equation}
t_\mathrm{cond}^{-1}\equiv\frac{1}{a}\left(\frac{da}{dt}\right)=s_\mathrm{a}V_0\left(\frac{\kb\tg}{2\pi m_\mathrm{atom}}\right)^{1/2}\left(1-\frac{1}{\mathcal{S}}\right)\frac{n_\mathrm{atom}}{a}, \label{eq:cond}
\end{equation}
where $n_\mathrm{atom}$ is the number density of impining atoms.
For the case of $\mathcal{S}=1.1$, typical condensation timescale of silicon atoms is about
\begin{eqnarray}
t_\mathrm{cond}\simeq \frac{1.6\times10^2}{s_\mathrm{a}}~\mathrm{yr}\left(\frac{n_\mathrm{Si}}{1.8\times10^{6}\ \mathrm{cm}^{-3}}\right)^{-1} \nonumber\\
\left(\frac{a}{3\ \mu\mathrm{m}}\right)\left(\frac{\tg}{1500\ \mathrm{K}}\right)^{-1/2}, \label{eq:tacc}
\end{eqnarray}
where $n_\mathrm{Si}=A_\mathrm{Si}\nh$ is the number density of silicon atoms.
We have assumed $\nh=5\times10^{10}\ \mathrm{cm}^{-3}$, which corresponds to the gas density at 0.4 pc away from the black hole with $10^8~M_\odot$ (see Equations (3) and (4) in \citet{ichi17b}). Since Equation (\ref{eq:tacc}) assumes that all silicon atoms are in the gas phase, this timescale should be regarded as the lower limit.

Another possibility of grain growth is coagulation, or the hit-and-stick growth of dust grains. 
The coagulation timescale of dust grains can be evaluated by
\begin{equation}
\tcoag\simeq\frac{1}{s_\mathrm{c} n_\mathrm{gr}\sigma_\mathrm{col}\Delta v},
\end{equation}
where $s_\mathrm{c}$ is the sticking probability, $n_\mathrm{gr}$ is the number density of dust grains, $\sigma_\mathrm{col}$ is the collision cross-section, and $\Delta v$ is the relative velocity. Using the dust-to-gas mass ratio $f$, the number density of grains is estimated to be $n_\mathrm{gr}=f n_\mathrm{H}m_\mathrm{H}/m_\mathrm{gr}$.
The relative velocity of grains determined by their Brownian motion is given by $\Delta v=(8 k_\mathrm{B}T/\pi \mu_\mathrm{gr})^{1/2}$, where $\mu_\mathrm{gr}$ is the reduced mass of grains. Suppose two grains are identical and each grain has the mass $m_\mathrm{gr}$, $\Delta v=(16 k_\mathrm{B}T/\pi m_\mathrm{gr})^{1/2}$. If each grain has radius $a$, the collision cross-section is $\sigma_\mathrm{col}=\pi(2a)^2=4\pi{a}^2$. By assuming homogeneous spherical grains, we obtain
\begin{eqnarray}
    \tcoag&\simeq&\frac{2.6\times10^{5}}{s_\mathrm{c}} \mathrm{yr} \left(\frac{f}{0.01}\right)^{-1}
    \left(\frac{n_\mathrm{H}}{5\times10^{10}\ \mathrm{cm}^{-3}}\right)^{-1}\nonumber\\
    &&
    \left(\frac{a}{3\ \mu\mathrm{m}}\right)^{5/2} 
    \left(\frac{\rho_s}{3.5\ \mathrm{g}\ \mathrm{cm}^{-3}}\right)^{3/2}
    \left(\frac{T}{1500\ \mathrm{K}}\right)^{-1/2},
\end{eqnarray}
where $\rho_s$ is the material density of dust grains. Except for very small grains (with a radius less than 0.02 $\mu$m for $T=1500$ K), the collision velocity is slower than the critical fragmentation velocity of silicate grains ($\sim \mathrm{m}~\mathrm{s}^{-1}$) \citep{C93}. 
Therefore, it is reasonable to assume the sticking probability of micron-sized grains $s_\mathrm{c}\simeq1$.

As a result, as long as the gas phase is sufficiently supersaturated, which might be expected near the sublimation radius \citep[e.g.,][]{ros13}, condensation may dominates grain growth.
Observations suggest the typical AGN lifetime of 10--100 Myr
\citep[e.g.,][]{mar04} and simulations \citep[e.g.,][]{hop06}
\citep[see also discussions in][]{ina18}. 
Hence, grain growth can potentially form micron-sized grains at the inner region of dusty torus. Even if the condensation is not efficient, coagulation may form large grains within the lifetime of AGN as long as the torus structure is static.

\subsection{Implications for the Silicate Feature of AGNs in the MIR}
\begin{figure}[t]
\begin{center}
\includegraphics[height=6.0cm]{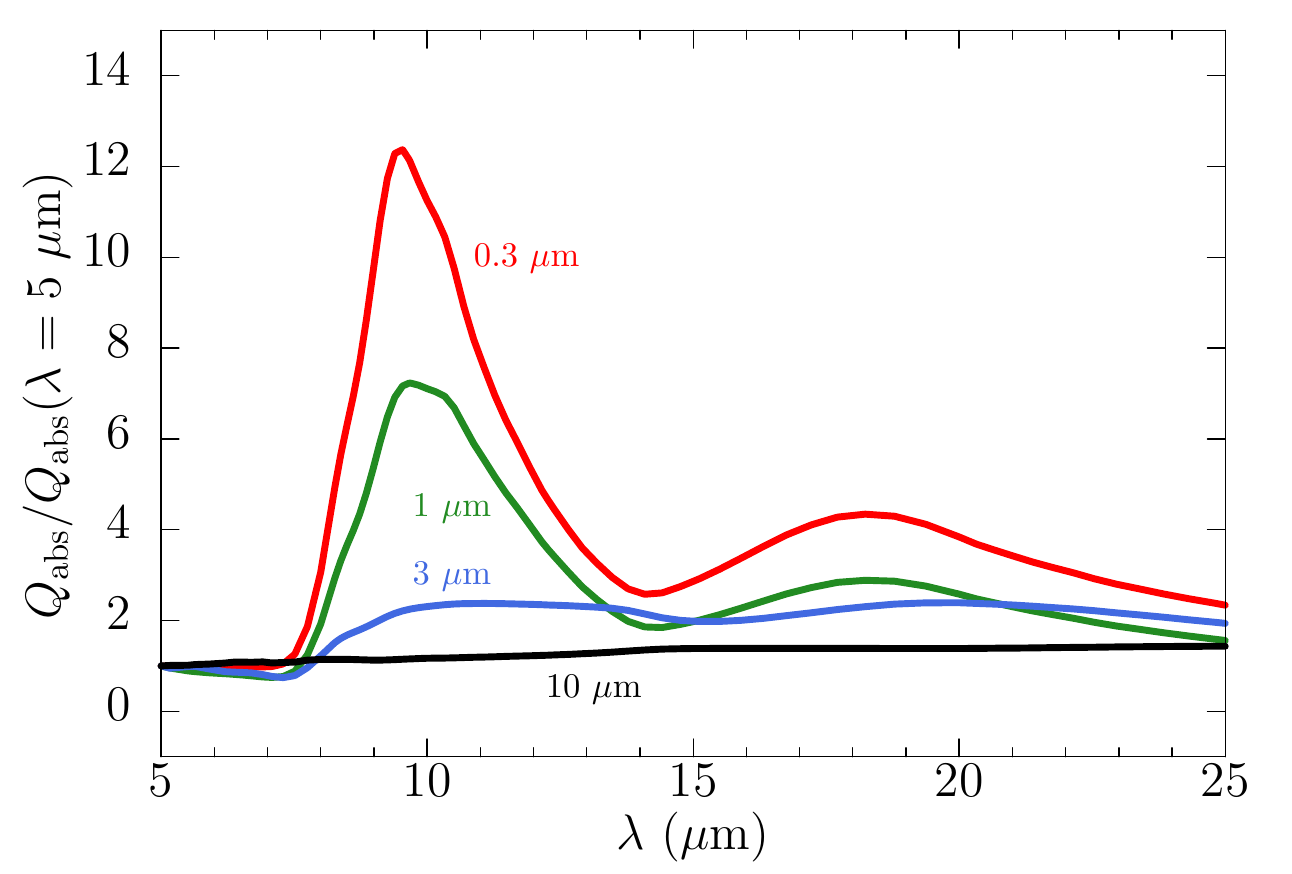}
\caption{Emission efficiency of silicate particles with various radii. The efficiency is normalized at $\lambda=5\ \mu$m. Refractive indices are astronomical silicate.}
\label{fig:silfeat}
\end{center}
\end{figure}

%
%
In Section \ref{sec:results}, 
we showed that small dust grains are more readily destroyed due to drift-induced kinetic sputtering. 
One key observational expectation is 
the change in strength of the silicate feature,
which can be observed in the MIR band.

Figure \ref{fig:silfeat} presents the emission efficiency of silicate particles as a function of 
wavelength for various dust radii. 
For particles smaller than $0.3\ \mu$m, the silicate feature at $\sim10$ and $\sim18$~$\mu$m is prominent.
In contrast, the silicate feature is attenuated if a dust grain is larger than $1.6~\mu$m$~\approx 9.8~\mu\mathrm{m}/2\pi$, or the size parameter exceeds unity for the silicate feature wavelength. When the size parameter exceeds unity, a dust grain with a moderate refractive index ($|m-1|\sim 1$, where $m$ is the complex refractive index) becomes optically thick for the radiation; in this case, the emission approaches that of a perfect blackbody. 
Therefore, the removal of small grains and/or the prevalence of large grains makes the silicate feature less prominent. 

%
%
The recent high spatial resolution MIR interferometric observations of the polar dust region often reveal a weak silicate feature \citep[e.g.,][]{hon12,bur13}
as well as SED decomposition of the dusty polar region \citep{lyu18}; these observations are consistent with our results suggesting that only large-sized dust grains can survive in such a region. 

\subsection{Silicate Feature among Type-1 AGN, Type-2 AGN, and ULIRGs as Buried AGN}
%
%

Drift-induced sputtering can be expected when the grains are exposed to the AGN radiation.
Hence, this process may also explain the
observed differences in the silicate features of type-1 and type-2 AGNs.

Conventionally, 
for type-1 (face-on) AGN, one would expect that the silicate feature
would be observed in the emission because the
heated silicate dust on the surface of the inner dusty disk can be observed directly, 
i.e., allowing for direct detection of the
silicate features in the emission. 
In contrast, for type-2 (edge-on) AGN,
the silicate feature would be visible
in absorption spectra, because the external dusty disk wall would block the inner dust emission.
Although the silicate feature shows a wide range of variety both in type-1 and type-2 AGN observationally  \citep[][]{lev07,wu09,haz15},
it is known that 
averaged type-1 (face-on) AGN spectra 
exhibit weak silicate emission,
unlike averaged type-2 (edge-on) AGN spectra that exhibit deeper silicate absorption \citep[e.g.,][]{H07}.

Such observational issues may be circumvented once
the smaller grains in the central few pc are disrupted under AGN irradiation. Here, both the dust in the dusty outflow region and the surface of the dusty disk producing a dust continuum with silicate-free features, as shown in Figure~\ref{fig:silfeat}, 
can be observed directly for type-1 AGN.
In contrast, the origin of the moderated level of absorption in type-2 AGN would be a combination of 1) the contribution of silicate absorption from the dusty disk, where the small-sized dust grains are still alive, and 2) the contribution of silicate feature-free emission from the partially visible dusty outflow region with a size of a few pc \citep[e.g.,][]{hon13}. 

If the above discussion is correct, 
it also naturally explains the dichotomy of silicate features between type-2 AGN and ultra-luminous infrared galaxies (ULIRGs) with a deeper silicate feature \citep[e.g.,][]{H07}. Because most of the ULIRGs are considered to be buried AGN obscured by dust in all directions \citep[e.g.,][]{ima10,ich14}, the dusty outflow region is also highly obscured even if it exists. Thus, ULIRGs have a deeper silicate absorption feature, compared with those of type-2 AGN.

\subsection{Comparison of Our Results with Previous Dust Studies of AGNs}

It is helpful to compare our results
with previous observational and theoretical
studies of dust properties in AGNs.
\cite{ait85} first reported that a small-sized 
grain with $a<0.01$~$\mu$m would be
easily destroyed in AGNs, and only
larger grains with $a>0.01$~$\mu$m
could survive due to the longer
lifetime.
\cite{lao93} also found that a
dust grain size of $>3$~$\mu$m is
preferable for AGNs with a deficit
of silicate features, and \cite{lao93} also discussed
that the optically-thick dusts also strongly reduces the amplitude of the silicate feature \citep[see Figure~16 in][]{lao93}.

\cite{bas18} further discussed
how the dust sublimation area depends
on the dust composition and the grain size; this 
suggests that the silicate grain
size distribution will be shifted to a
larger size, for a grain size distribution of up to $a\simeq1$~$\mu$m.
Our results also indicate that
a much larger grain size with $a>1$~$\mu$m would be preferable in the polar dust region at a few pc distance from the central engine; otherwise, the dust would be completely evaporated at that distance. The advantage of our result
is that it is no longer necessary to consider the dust composition issue, i.e., how only purely carbon dust can be lifted up to the polar dust region without silicate dust contamination.

\section{Conclusion} \label{sec:conclusion}
We have studied the grain dynamics and destruction around a geometrically thin AGN dusty disk by considering the radiation pressure from the AGN and dusty disk IR emission. In this paper, we have proposed a new dust destruction mechanism at the AGN environment, drift-induced sputtering, in which highly accelerated grains due to AGN radiation pressure are destroyed by kinetic sputtering (Figures \ref{fig:trac1}, \ref{fig:trac2}, \ref{fig:trac3}, and \ref{fig:trac4}). As a result, smaller grains ($\lesssim\mu$m) are rapidly destroyed by kinetic sputtering, whereas larger grains ($\gtrsim\mu$m) tend to survive over a longer timescale, leading to longer blown-out distance for larger grains (Figure \ref{fig:rdead}). Prevalence of micron-sized or more larger grains can naturally explain a lack of silicate features in MIR emission (Figure \ref{fig:silfeat}), typically observed for type-1 AGN as well as polar dust emission.

The hypersonic drift occurs when the Coulomb drag force is weaker than radiation pressure (Figure \ref{fig:fdrag}). For lower gas density and/or higher gas temperature, Coulomb drag tends to be weaker than radiation pressure (Figure \ref{fig:cond}). 
In addition, the Coulomb drag force depends also on grain radius, and larger grains experience weaker Coulomb drag (Figure \ref{fig:fdrag}d). For example, when $\nh=10^2~\unitnum$ and $\tg=10^4$ K at 1 pc away from the nucleus, the hypersonic drift is halted by the Coulomb drag force for grains smaller than 0.1 $\mu$m. Although these small grains are not destroyed by sputtering, Coulomb explosion may eliminate these grains \citep{RT20a}.

We have also investigated the role of dusty disk radiation pressure on dust grain trajectories. Dusty disk radiation pressure can levitate dust grains when the dusty disk luminosity is as large as or higher than the AGN luminosity. Since such a luminous dusty disk seems to be unrealistic, the role of dusty disk emission for the levitation of grains is rather limited.

Finally, we have shown
the large ($>1$~\micron) grains can be
produced in the inner region of the AGN
dusty disk within a timescale shorter than the typical AGN lifetime, indicating that
the large grains are pervasive throughout the AGN dusty disk.

\acknowledgments
We thank the anonymous referee for useful comments which significantly improve the quality of this paper.
R.T. would like to thank Joseph Weingartner and Takaya Nozawa for kind cooperation of our code validation. 
We also thank Mitsuru Kokubo, Thiem Hoang, and Taiki Kawamuro for fruitful discussions. 
R.T. was supported by a Research Fellowship for Young Scientists from the Japan Society for the Promotion of Science (JSPS) (JP17J02411). This work was also supported by JSPS KAKENHI Grant Numbers JP19H05068 (R.T.) and JP18K13584 (K.I.).
This work was supported by the Program for Establishing a Consortium
for the Development of Human Resources in Science
and Technology, Japan Science and Technology Agency (JST).

\appendix 
\section{Grain Charging Rate} \label{sec:jrate}
\begin{figure*}[tbp]
\begin{center}
\includegraphics[height=6.0cm,keepaspectratio]{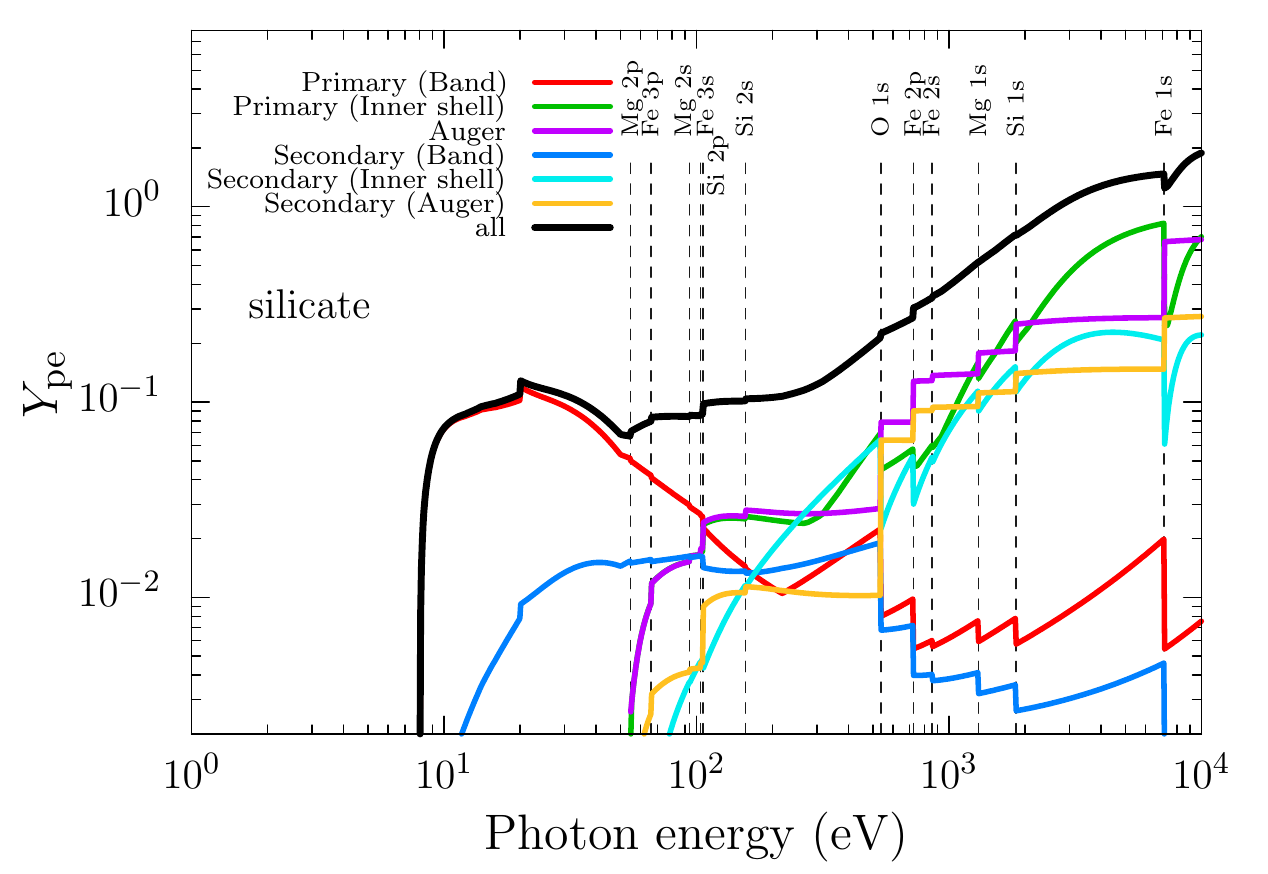}
\includegraphics[height=6.0cm,keepaspectratio]{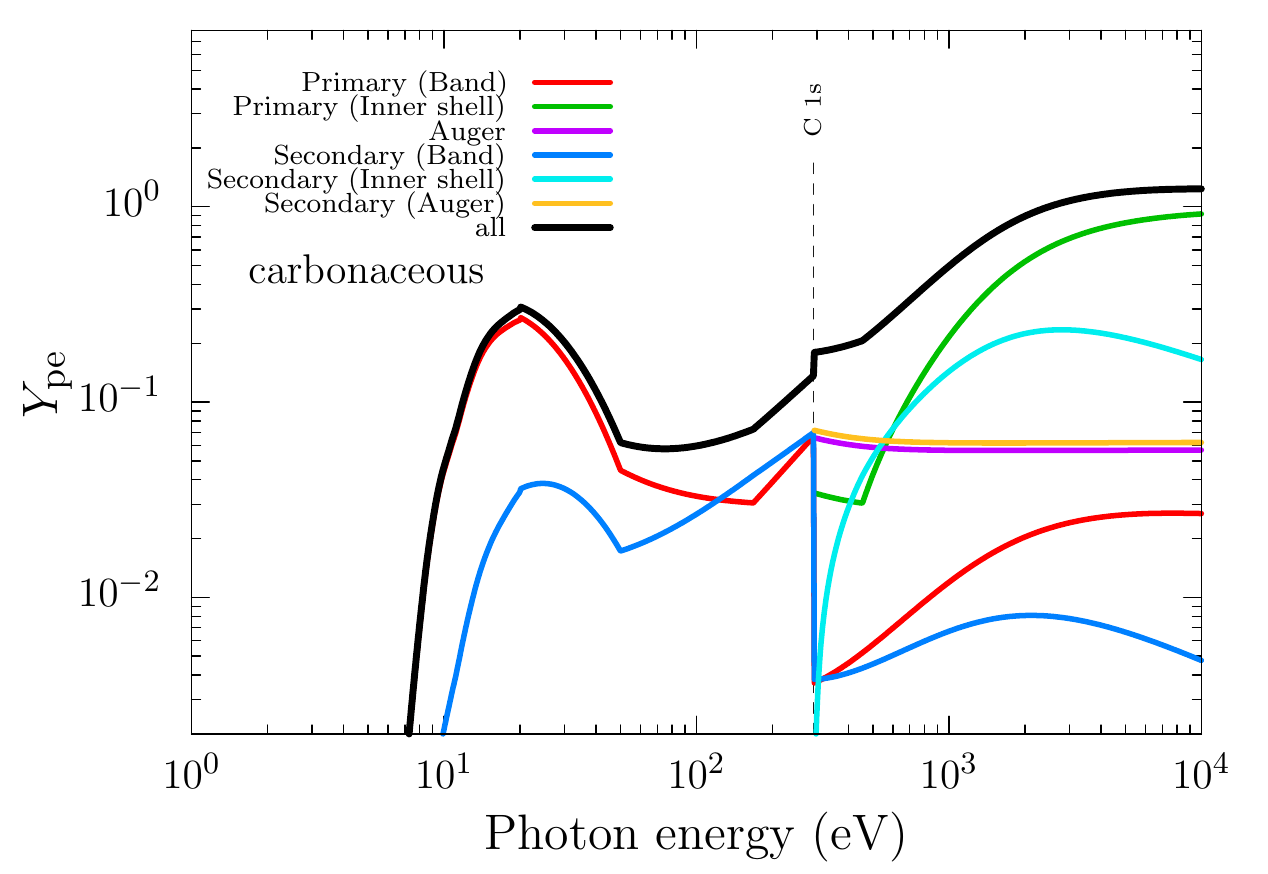}
\includegraphics[height=6.0cm,keepaspectratio]{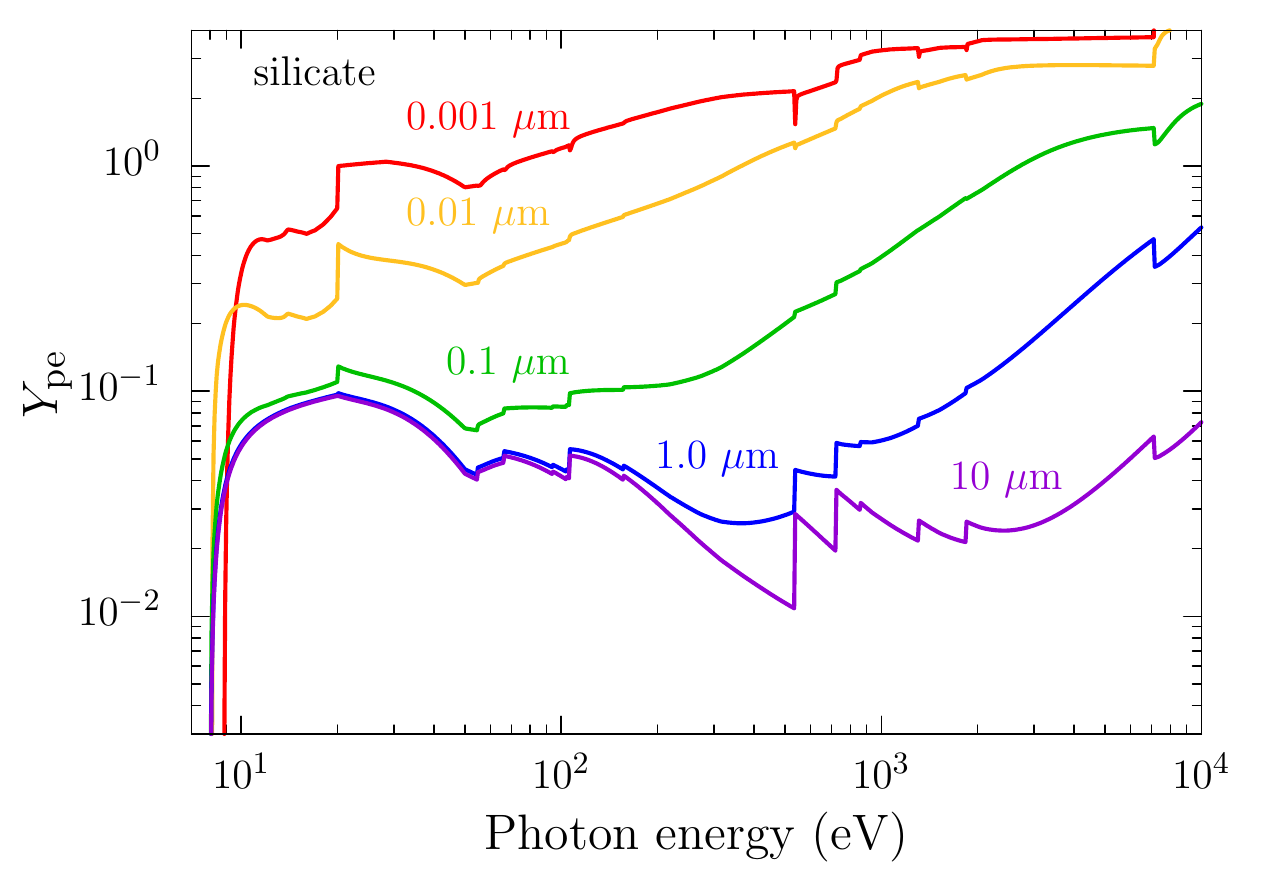}
\includegraphics[height=6.0cm,keepaspectratio]{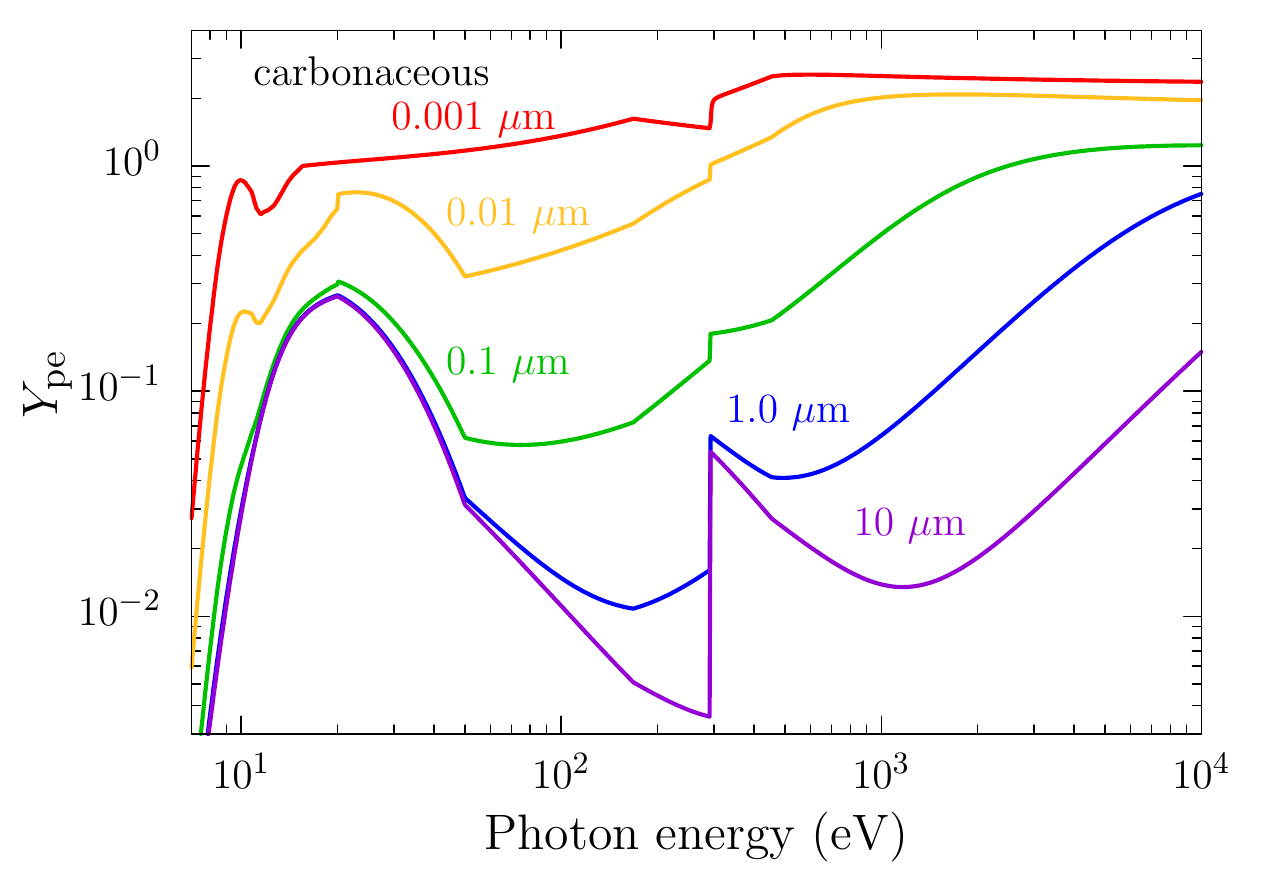}
\caption{Photoelectric yield $\ype$ used in this paper. Left and right panels show silicate and carbonaceous grains, respectively. (Top): $\ype$ for uncharged grains with $a=0.1~\mu\mathrm{m}$. Contribution from primary, secondary, and auger transitions are shown. Ionization energies of the inner shells of atomic species composing the grain are indicated by vertical dashed lines. (Bottom): Grain size dependence on $\ype$.}
\label{fig:ype}
\end{center}
\end{figure*}

The photoelectric emission rate is estimated by \citep{WD01}
\begin{equation}
\jpe=\sigd\int_0^\infty d\nu \frac{cu_\nu}{h\nu}\qabs(\nu)\ype(\nu),
\end{equation}
where $\qabs$ is the photon absorption efficiency and $\ype$ is the photoelectric yield. 
In this paper, we compute $\ype$ based on a model described in \citet{WDB06}. Namely, we consider photoelectrons of primary and secondary origins as well as those emanated from auger transitions. The work functions of silicate and graphite are estimated to be about 8 eV and 4.4 eV \citep{WD01}. We take the photoionization cross sections from \citet{V95} and \citet{V96}, and auger transitions data are taken from \citet{D96}.
Figure \ref{fig:ype} shows the $\ype$-values for silicate and carbonaceous grains. Photoelectric yield becomes higher for smaller particles because of small particle effect \citep{W73, D78}. A small particle is expected to show higher photoelectric yield because an incident photon is likely absorbed nearer the surface than if it is absorbed in a slab, or the bulk material \citep{W73}. As a result, photoelectron can more readily escape from the grain than the case of the bulk material. We ignore the surface curvature effect on the yield \citep{K16}, which may elevate the yield-value for very small grains, such as nm-sized grains.

The collisional charging rate is given by \citep{D87}
\begin{equation}
\je=\sigd  n_e s_e \left(\frac{8\kb\tg}{\pi m_e}\right)^\frac{1}{2}\tilde{J}\left(\tau_e=\frac{a\kb\tg}{q_e^2},\nu_e=\frac{e\zd}{q_e}\right), \label{eq:je}
\end{equation}
where $s_e$ is the sticking probability of colliding electrons, and $m_e$ and $q_e=-e$ are mass and charge of the electron. $\tilde{J}$ is the correction term arising from Coulomb interaction between (i) an impacting particle and net-grain charge and (ii) an impacting particle and the image charge resulting from the polarization of grain induced by the Coulomb field of an impacting particle. 
The sticking probability is $s_e=0.5(1-e^{-a/l_e})$, where $l_e$ is the electron escaping length given in Equation (12) and (13) in \citet{WDB06}.
The correction of collision cross section due to Coulomb interaction, $\tilde{J}(\tau,\nu)$, is estimated by \citep{D87},
\begin{eqnarray}
\tilde{J}(\tau,\nu=0)&=&1+\left(\frac{\pi}{2\tau}\right)^{1/2},\label{eq:jtil1}\\
\tilde{J}(\tau,\nu<0)&\approx&\left[1-\frac{\nu}{\tau}\right]\left[1+\left(\frac{2}{\tau-2\nu}\right)^{1/2}\right],\label{eq:jtil2}\\
\tilde{J}(\tau,\nu>0)&\approx&[1+(4\tau+3\nu)^{-1/2}]^2\exp(-\theta_\nu/\tau),\label{eq:jtil3}
\end{eqnarray}
where $\theta_\nu=\nu/(1+\nu^{-1/2})$. Equation (\ref{eq:jtil1}) is an exact relation, while Equations (\ref{eq:jtil2} and \ref{eq:jtil3}) are approximate formulae, which agree with numerical results to an accuracy of $\pm$5\% for $10^{-3}<\tau<\infty$ and $\nu\le-1$ and $\pm4$\% for $\nu\ge1$, respectively.
$\theta_\nu$ in Equation (\ref{eq:jtil3}) is also approximate formula, which is accurate within 0.7\% for $1\le\nu<\infty$.

For the secondary gas emission rate, $\jsec$, is also computed from Equation (\ref{eq:je}) by using the secondary emission probability $\delta_e$ instead of $s_e$. We estimate $\delta_e$ by using a model described in \citet[][see their Equations 14-16 and A7-A11]{D79}.
Similarly, the collisional charing due to ions is
\begin{equation}
\jion=\sigd n_\mathrm{H} \sum_i A_i s_i\left(\frac{8\kb\tg}{\pi m_i}\right)^\frac{1}{2} \tilde{J}\left(\tau_i=\frac{a\kb\tg}{q_i^2},\nu_i=\frac{e\zd}{q_i}\right),
\end{equation}
where $s_i$ is the sticking probability, $m_i$ and $z_i$ are mass and charge of ion, and the summation $i$ is taken over various atomic species. In this paper, we assume $s_i=1$.

\section{Sputtering Yield model} \label{sec:ysp}
\begin{figure*}[t]
\begin{center}
\includegraphics[height=6.0cm,keepaspectratio]{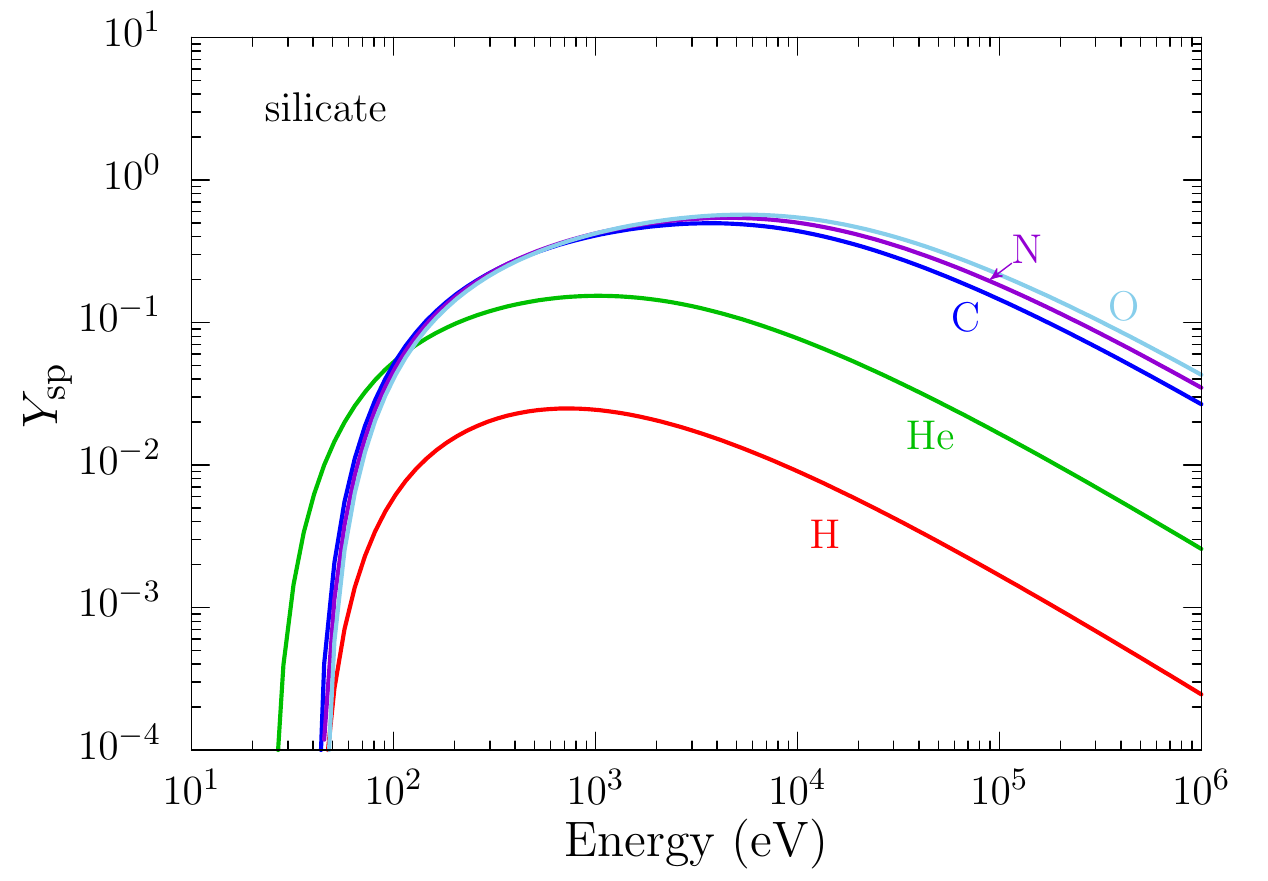}
\includegraphics[height=6.0cm,keepaspectratio]{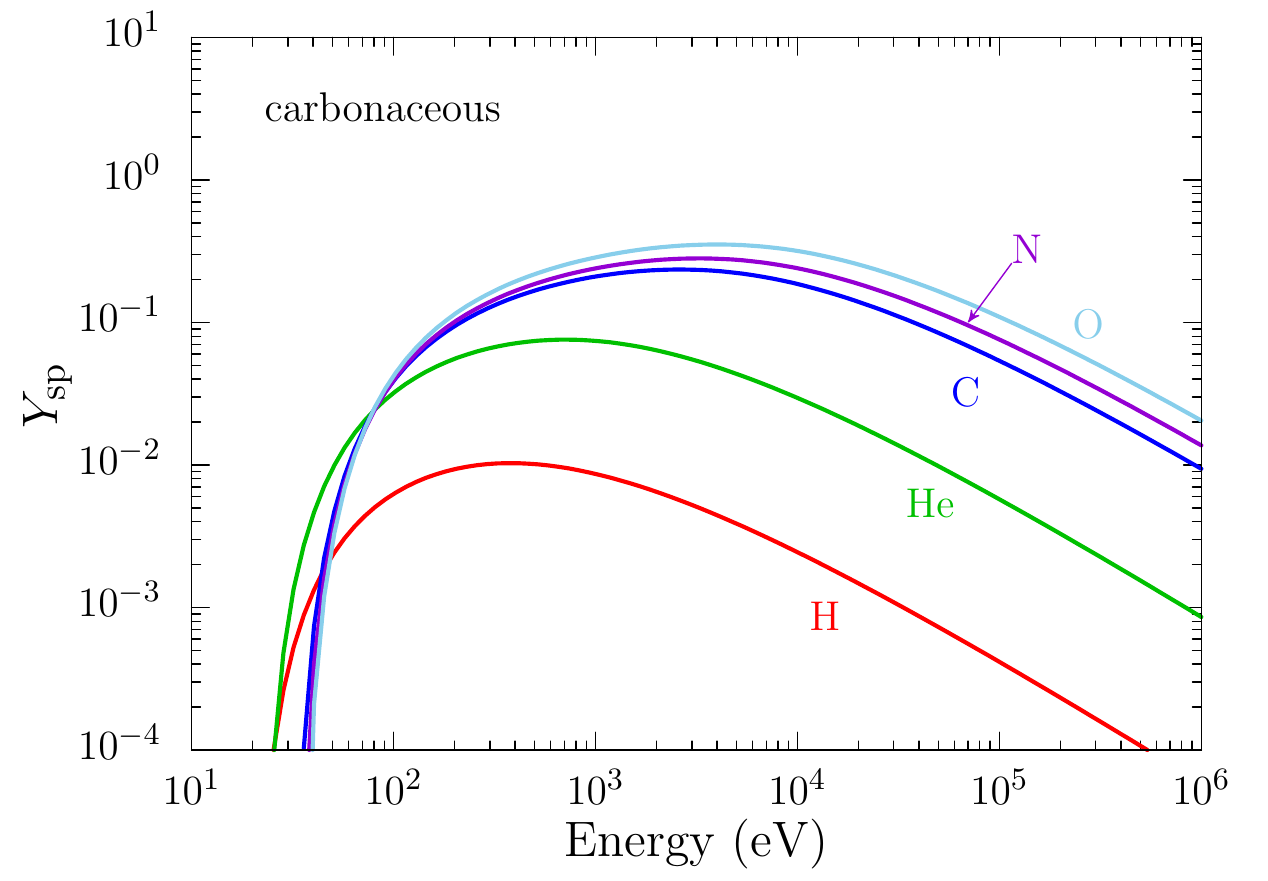}
\caption{Sputtering yield adopted in this study. Left and right panels show sputtering yield of silicate and carbonaceous grains, respectively. Different lines correspond to different impacting atoms.}
\label{fig:ysp}
\end{center}
\end{figure*}

The sputtering yield is calculated based on the model described in \citet{N06}.
Figure \ref{fig:ysp} shows the sputtering yield of silicate and carbonaceous grains. The sputtering rate was based on bulk material measurements. Note that the sputtering yield in the high-energy regime, where the penetration length of an incident particle becomes comparable to or larger than the grain size, is still unclear. \cite{J98} reported that the sputtering yield increased in such a regime, although more detailed studies are necessary to quantify this effect.

\bibliography{cite}

\end{document}